\documentclass[global,floatfix,twocolumn,referee,pre,aps]{revtex4-1}
\usepackage{graphicx}
\usepackage{bm}

\usepackage[ngerman,british]{babel}

\usepackage{amsmath,amssymb,amsthm,dsfont,mathrsfs,bbm}

\usepackage{array}
\usepackage{multirow}	
\usepackage{booktabs}						
\usepackage{dcolumn} 						

\usepackage{graphicx}
\usepackage{subcaption}  
\usepackage{wrapfig}
\usepackage{savesym}

\usepackage{tikz}
\usetikzlibrary{matrix,shapes,arrows,positioning,chains}
\usepackage{nomencl}

\newlength{\tmplen}

\savesymbol{a}
\savesymbol{b}
\savesymbol{d}
\savesymbol{e}
\savesymbol{k}
\savesymbol{l}
\savesymbol{H}
\savesymbol{L}
\savesymbol{p}
\savesymbol{r}
\savesymbol{s}
\savesymbol{S}
\savesymbol{F}
\savesymbol{t}
\savesymbol{w}
\savesymbol{th}
\savesymbol{div}
\savesymbol{tr}
\savesymbol{Re}
\savesymbol{Im}
\savesymbol{AA}
\savesymbol{SS}
\savesymbol{Si}

\newcommand{\e}{\varepsilon}
\newcommand{\d}{\delta}

\newcommand{\G}{\Gamma}
\newcommand{\l}{\lambda}

\newcommand{\r}{\rho}
\newcommand{\s}{\sigma}
\newcommand{\t}{\tau}

\newcommand{\dt}{{\Delta t}}
\newcommand{\dx}{{\Delta x}}

\newcommand{\FF}{{\mathcal F}}
\newcommand{\HH}{{\mathcal H}}
\newcommand{\OO}{{\mathcal O}}

\newcommand{\cov}{{\nabla}}

\newcommand{\del}{\partial}
\newcommand{\delvar}{{\overline \partial}}

\newcommand{\pdiff}[2]{\frac{\partial#1}{\partial #2}}

\newcommand{\D}{\displaystyle}
\newcommand{\T}{\textstyle}

\makenomenclature

\begin{document}

\preprint{APS/123-QED}

\title{Fluid structure interaction with curved space lattice Boltzmann}
\author{Kyriakos Flouris}
\author{Miller Mendoza Jimenez}
\author{Gautam Munglani}
\author{Falk K. Wittel}
\author{Jens-Daniel Debus}
\author{Hans J. Herrmann}
\affiliation{%
Computational Physics for Engineering Materials, Institute for Building Materials, Wolfgang-Pauli-Str. 27, HIT, CH-8093, Zurich  Switzerland
}

\begin{abstract}

We present a novel method for fluid structure interaction (FSI) simulations where an original 2$^{nd}$-order curved space lattice Boltzmann fluid solver (LBM) is coupled to a finite element method (FEM) for thin shells. The LBM  can work independently on a standard lattice in curved coordinates without the need for interpolation, re-meshing or an immersed boundary. The LBM distribution functions are transformed dynamically under coordinate change. In addition, force and momentum can be calculated on the nodes exactly in any geometry.  Furthermore, the FEM shell is a complete numerical tool with implementations such as growth, self-contact and strong external forces. We show resolution convergent error for standard tests under metric deformation.  Mass and volume conservation, momentum transfer, boundary-slip and pressure maintenance are verified through specific examples. Additionally, a brief deformation stability analysis is carried out. Next, we study the interaction of a square fluid flow channel to a deformable shell. Finally, we simulate a flag at moderate Reynolds number, air flow channel. The scheme is limited to small deformations of $\mathcal{O}(10\%)$ relative to domain size, by improving its stability the method can be naturally extended to multiple applications without further implementations.
\end{abstract}

\maketitle
\section{Introduction}

Fluid flow is commonly confined through pipes, between sheets, or within some other arbitrary solid surface. A compelling challenge for simulations is the situation when these confining surfaces strongly deform. Abundant examples can be found in biology \cite{fsi_review_heart, review_heart2},  and technology \cite{fsi_book}, on both the macro and micro scales, growth or external forces. The confining cavity constrains the fluid flow and vice versa. In the case of large deformations, possibly caused by the fluid flow, the surface wall can take a complicated shape with high local curvatures \cite{high_local_curvature}  or even develop localized ridges or kinks. In such situations inertial fluid effects can exert large forces and generate centrifugal instabilities. The physical consequences of this feedback, if properly understood, can bring a new perspective to biological morphogenesis or biomorphic technologies.  Important applications include temperature control of fluids \cite{temp_control_fluids} or chemical reactions on a surface \cite{surface_chemistry}; like combustion and metabolism which require mass transport by the fluid. In summary, instabilities and conservation laws that arise from the coupling between the fluid motion and the wall structure can cause various physical effects such as buckling, wriggling, clogging, crumpling, necking, pinching and cavitation. These can be represented as a moving boundary problem which poses considerable numerical challenges, strongly suggesting the need to explore new methods and techniques. 

Fluid structure interaction (FSI) simulations allow for an efficient advance in industrial design and optimization \cite{FSI_engine}. Furthermore, in bio-science and biotechnology the ever increasing complexity of study cases and implementations \cite{FSI_heart, FSI_vesels, fsi_review_heart}, demands the use of accurate and efficient algorithms in simulating such configurations.  Numerous fluid structure interaction implementation methods exist, utilizing various techniques with different outcomes. More 'traditional' methods can include finite element solvers \cite{finite_elements_book}, which have been implemented by commercial codes such as ADINA or COMSOL with extreme success. When it comes to complicated geometries, most of the existing methods are designed to work on a Cartesian grid. A boundary change can either be accommodated by re-meshing (conforming mesh methods)  \cite{review_fsi_types}  or  by an immersed boundary  (non-conforming mesh methods) \cite{immersed, immersed2}. Re-meshing can be very computationally expensive especially for a moving body or boundary. Whereas an immersed boundary solver can be problem specific and the computational cells are cut and solutions interpolated, because a  displaced boundary would not align with the grid.Additionally, the above methods have been used to couple LBM to FEM \cite{fsi_lbm_fem1,fsi_lbm_fem2,fsi_lbm_fem3} with equivalent limitations.


We propose a description for FSI simulations of various physical scenarios using the lattice Boltzmann method (LBM) \cite{LBM_review, LBM_initial} coupled to a Finite Element Method (FEM). To achieve this we implement a generalization of the LBM in curved space proposed by Mendoza et al. \cite{miller_lks}, adapted to second order, to simulate the fluid as well as a FEM created by Vetter et al \cite{shellcode, roman_thesis} to simulate a shell boundary. The LBM can solve the Navier-Stokes equations (NSE) in arbitrary curvilinear coordinates and the FEM shell is very robust during large deforming. 

Standard LBMs are restricted to regular grids.  To overcome this limitation off-lattice Boltzmann methods have been developed. These include finite-volume  \cite{OL-finitevolume, OL-finitevolume2}, finite-element \cite{OL-finiteelement}, and finite-difference \cite{OL-finitediff, OL-finitediff2} methods. All of the above schemes are limited by interpolation-supplemented unstructured grids. 

The elegance of our approach is the ability for the LBM  to work independently on a standard lattice in curved coordinates without the need for interpolation, re-meshing or an immersed boundary. In addition, force and momentum can be calculated on the nodes exactly in any geometry, which  can then be transformed back to Cartesian coordinates.  
 
 Simultaneously the FEM shell, apart from stability advantages, utilizes a continuum description of thin elastic objects with inherent $C^2$ differentiability. This allows for very accurate position and velocity calculation.  Furthermore, this formulation would open up the possibility of exploiting the proven advantages of Lattice Boltzmann methods, namely computational efficiency and easy handling of complex geometries. Equally as important, the FEM shell is a complete numerical tool with implementations such as anisotropic or differential growth, self-contact, contact between different flexible shapes, spatial constraints and strong external forces. Therefore,  utilizing differential geometry and a complete FEM shell solver, a clean, general solver for FSI can be produced.

Firstly we introduce the LBM, followed by the basics of the FEM and especially the subdivision surface. Consequently the coupling procedure is described in detail including the transformation of fields. A validation and tests section precedes a simulations section where the main implementations are presented. The paper finishes with a summary and conclusions. Nomenclature and an appendix section can be found at the end. 

\nomenclature{LBM}{ lattice Boltzmann method}%
\nomenclature{FEM}{ finite elements method}%

\section{Lattice Boltzmann in curved space}
The LBM was developed to simulate fluids by means of simple arithmetic operations instead of directly solving the macroscopic equations of continuum fluid mechanics (i.e. NSE). This is achieved by performing simple arithmetic operations based on the Boltzmann equation, derived from the kinetic theory of gases and it defines the microscopic motion of fluid particles \cite{lbm_review_succi}. Due to its generality, the LBM has also been applied to other, similar, differential equations such a quantum mechanics \cite{succi_qlbm} and relativistic flows \cite{miller_rlbm}. Furthermore, an LBM has been developed to operate in general Riemannian manifolds \cite{debus_curved}, which enables the simulation of flows in arbitrary geometries. The main idea behind this method is to solve all the relevant equations in the basis of the tangent space of a curved manifold. The solution, once transformed to Cartesian coordinates (Euclidean space), is independent of the geometry as long as there is no intrinsic curvature of the manifold (trivial Ricci scalar and Ricci tensor). Please refer to appendix~\ref{app:riemannian} for a brief description of the relevant differential geometry.

\begin{figure}
\includegraphics[width=\columnwidth, height=2.5in]{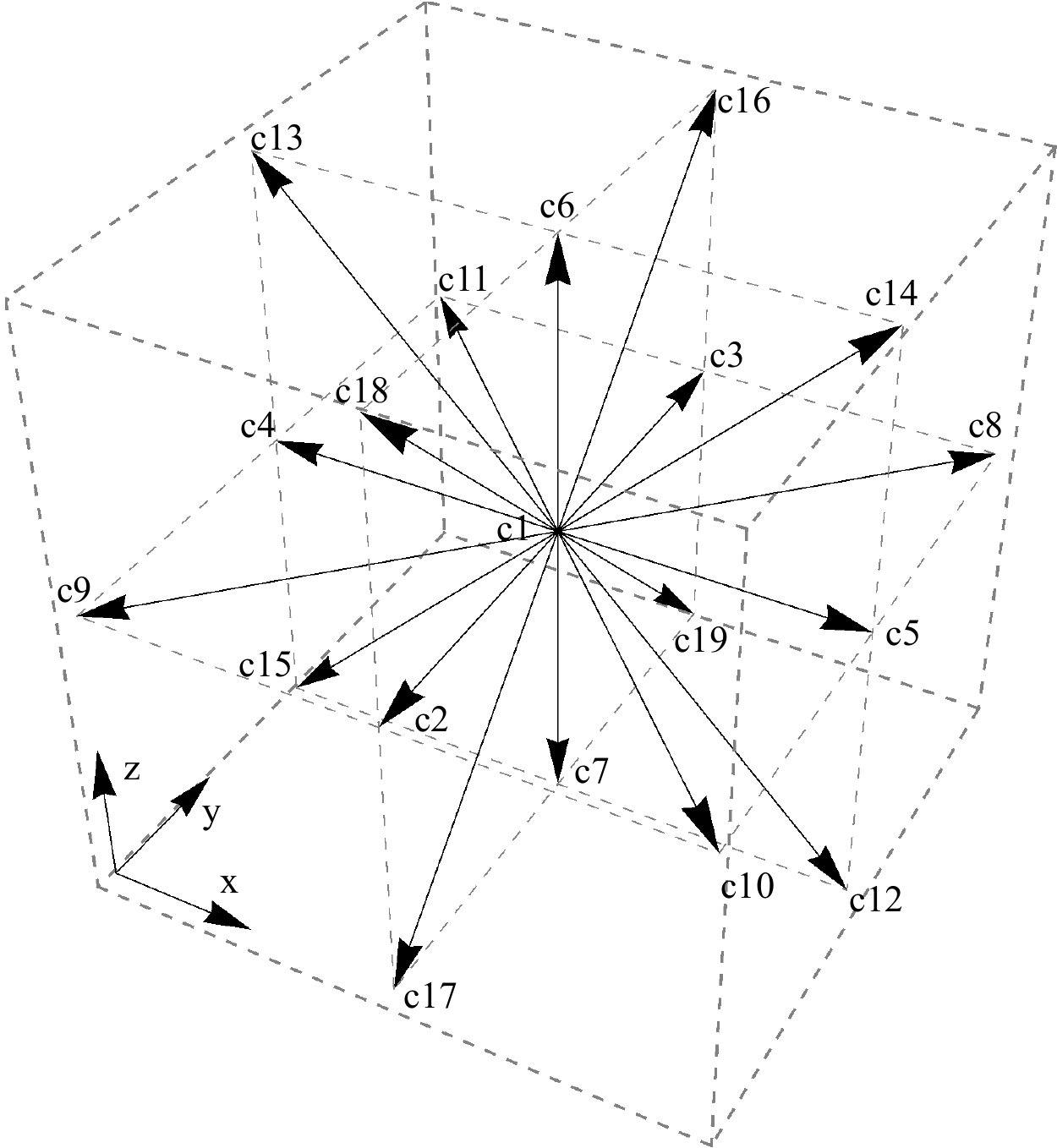}
\caption{\textbf{LBM velocity vectors, $\mathbf{D3Q19}$.} The 3D lattice Boltzmann velocity vectors at each lattice point are shown. }
\label{fig:lb_velocities}
\end{figure}

We hereby propose a 2$^{nd}$-order curved space LBM as modified from Ref.~\cite{miller_campylotic}.  The method  implements a 2$^{nd}$-order Hermite expansion of the distribution functions with second nearest neighbors LBM velocity vectors ($D3Q19$) as shown in Fig.~ \ref{fig:lb_velocities}. Starting from the curved space LBM equation, defining the equilibrium distributions and the forcing term, one can show with a Chapman-Enskog expansion procedure that the NSE are recovered with error of $\mathcal{O}(\Delta t^2)$ ( for derivation see Appendix~\ref{app:chapmanenskog}). In this work the Latin indices run over the spatial dimensions and Einstein summation convection is used for repeated indices, unless otherwise stated.
 
The 2$^{nd}$-order curved-space LB equation is given by:
\begin{align}
\label{eq:CE-LB}
	f_\l(x + c_\l \dt, t + \dt) - f_\l(x, t)  = - \frac{1}{\t} \left( f_\l - f_\l^{\rm eq} \right) 	 + \dt \FF^*_\l
\end{align}
The equation describes the evolution of a discrete distribution function $\{ f_{\l}(x,t) \}_{\l=1}^Q$ on the lattice,  the $\lambda$ label corresponds to the velocity vector $c_\lambda$ ( for each LBM node, shown in Fig.~\ref{fig:lb_velocities}, $Q$ is the total number of these velocity vectors).

\nomenclature{$f_\l$}{ denotes the particle distribution function, depending on the local coordinate $x=(x^1,...,x^D)$}%
\nomenclature{$\lambda$}{label corresponds to the velocity vector $c_\lambda$}%
\nomenclature{$F^{eq}$}{ the equilibrium distribution function}%
\nomenclature{$\FF_\l$}{ the forcing term}%
\nomenclature{$\dt$}{ time step}%
\nomenclature{$\dx$}{ lattice spacing}%
\nomenclature{$c_\lambda$}{ velocity vector}
\nomenclature{$\t$}{ relaxation time}

In the kinetic theory of gases, $f=f (x,v,t) $ corresponds to the particle distribution function, depending on the local coordinate $x=(x^1,...,x^D)$, on the microscopic velocity $v=(v^1....,v^D)$ as well as on time $t$. $\tau$ is the relaxation time, $f^{eq}$ is the equilibrium distribution function and $\FF_\l$ is the forcing term (inertial forces on the manifold). In our case, the time step $\dt$ has the same value as the lattice spacing $\dx$.
Eq.~(\ref{eq:CE-LB}) converges to the covariant NSE at 2$^{nd}$ order in space and time,
\begin{align} 
	\del_t \r + \cov_i \left(\r u^i \right) = 0 + \OO(\dt^2), \notag\ \\
	\del_t \left(\r u^i\right) + \cov_j T^{ij} = 0 + \OO(\dt^2),
\end{align}
where the covariant derivative acting on a general vector $R^i$ is defined by $\cov_k (R^i) = \del_k (R^i) + \G^i_{kl} (R^l)$. $ \G^i_{jk}$ are the connection coefficients of the covariant derivative (or Christoffel symbols), please refer to Appendix~\ref{app:riemannian} for the definition and how it relates to the metric. $\rho$ and $u^i$ are the macroscopic density and velocities respectively:
\begin{align}
\label{eq:fields1}
\r &= \sum_\l f_\l \, \sqrt g = \sum_\l f^{\rm eq}_\l \, \sqrt g, \\
\label{eq:fields2}
\r u^i &= \sum_\l f_\l c_\l^i \, \sqrt g = \sum_\l f^{\rm eq}_\l c_\l^i \, \sqrt g. 
\end{align}
$T^{ij}$ denotes the energy-momentum tensor, which is composed of the free momentum-flux tensor $\Pi^{{\rm eq},ij}$ (equilibrium part), the viscous stress tensor $\s^{ij}$ (non-equilibrium part) and $g=det(g_{\mu\nu})$ (determinant of the metric tensor):
\begin{align} 
\label{eq:fields3}
T^{ij} &= \Pi^{{\rm eq},ij} - \s^{ij} \notag\ \\
&= \sum_\l f^{\rm eq}_\l c_\l^i c_\l^j \,  \sqrt g  + \left( 1 - \frac{1}{2\t}\right)   \sum_\l (f_\l - f^{\rm eq}_\l) c_\l^i c_\l^j \, \sqrt g.
\end{align}
\nomenclature{$\rho$}{  macroscopic density}%
\nomenclature{$u^i$}{  macroscopic velocities}%
\nomenclature{$T^{ij}$}{  energy-momentum tensor}%
\nomenclature{$\s^{ij}$}{ viscous stress tensor}%
\nomenclature{$g=det(g_{\mu\nu})$}{  determinant of the metric tensor }%
\nomenclature{$\Pi^{{\rm eq},ij}$}{  free momentum-flux tensor }%

The equilibrium distribution is expanded into Hermite polynomials in velocity space:
\begin{align}\label{eq:expansion-equilibrium-distribution}
	f_\l^{\rm eq} = \frac{w_\l}{\sqrt g} \sum_{n=0}^{3} \frac{1}{n!\,c_s^n}\, a_{(n)}^{{\rm eq},i_1,\ldots,i_n}\, \HH_{(n),\l}^{i_1,\ldots,i_n},
\end{align}
\nomenclature{$w_\l$}{   weight functions in the Hermite polynomials }%
$w_\l$ are the weight functions, described in detail in Appendix~\ref{sec:appendix_Hermite}.

In standard LB schemes, the forcing term is known to generate spurious artifacts at first order in $\Delta t$ \cite{discrete_lattice_effects}. A way to cancel this discrete lattice effects is to employ the trapezoidal rule for the time integration in the LB equation by using an improved forcing term \cite{Debus_thesis}, given by
\begin{align}
\FF^*_\l(x, t) := \FF_\l(x, t) 	+ \T\frac{1}{2} \left( \FF_\l(x + c_\l \dt, t ) - \FF_\l(x, t - \dt) \right),
\end{align}
where $\FF_\l$ is also expanded into Hermite polynomials:
\begin{align}	\label{eq:forcing2}
	\FF_\l 	&= \frac{w_\l}{\sqrt g} \sum_{n=0}^{2} \frac{1}{n!\,c_s^n}\, b_{(n)}^{i_1,\ldots,i_n}\, \HH_{(n)}^{i_1,\ldots,i_n}.
\end{align}
In order to match the NSE, the expansion coefficients must be chosen as follows:
\begin{align}
	b_{(0)} = A,  \qquad	b_{(1)}^{i} = B^i,  \qquad	b_{(2)}^{ij} = C^{ij} - c_s^2 \delta^{ij} A,
\end{align}
\nomenclature{$b_{(i)}$}{   expansion coefficients of $\FF_\l$ }%
where
\begin{align}
	\label{eq:forcing-A}
	A &= \sum_\l \FF_\l \,\sqrt g 	= - \G^i_{ij} \r u^j - \G^j_{ij} \r u^i, \\
	\label{eq:forcing-B}
	B^i &= \sum_\l \FF_\l c_\l^i \,\sqrt g \notag\ \\ 
	 &= - \G^k_{jk} T^{ij}  	- \G^i_{jk} T^{jk}   	- \G^j_{jk} T^{ki} 	+ F^{{\rm ext},i} , \\
	\label{eq:forcing-C}
	C^{ij} &= \sum_\l \FF_\l c_\l^i c_\l^j \,\sqrt g \notag\ \\ 
	&= c_s^2\, \delvar_k (\r u^i)\, \d^{jk}	+ c_s^2\, \delvar_k (\r u^j)\, \d^{ik}	+ c_s^2\, \delvar_k (\r u^k)\, \d^{ij} \notag\ \\
	& \ \ \ 
	- \theta\, \cov_k(\r u^i)\, g^{jk}	- \theta\, \cov_k(\r u^j)\, g^{ik}	- \theta\, \cov_k(\r u^k)\, g^{ij},
\end{align}
\nomenclature{$c_s$}{   lattice specific speed of sound }%
\nomenclature{$\theta$}{   normalized temperature  }%
where
\begin{align*}
 \delvar_k &= \del_k - \G^i_{ki}.  \notag\ 
\end{align*}
\nomenclature{$\G^i_{jk}$}{   connection coefficients of the covariant derivative  }%

The derivatives $\del_k (\r u^i)$ can be computed very accurately by using discrete isotropic lattice derivative operators please refer to the appendix \ref{app:derivative} for details.
		
In order to solve the LB equation numerically, the equation is typically split into a collision step
\begin{align}
 \label{eq:collision}
 f_\l^*(x, t) = - \frac{1}{\t} \left( f_\l(x,t) - f_\l^{\rm eq}(x,t) \right) \notag\ \\ 
	 + \dt\, \FF_\l(x,t) - \frac{\dt}{2} \FF_\l(x, t - \dt),
\end{align}
and a streaming step
\begin{align}
   \label{eq:streaming}
	f_\l(x, t + \dt) = f_\l^*(x - c_\l \dt, t) + \frac{\dt}{2} \FF_\l(x, t),
\end{align}
where $f^*$ denotes an auxiliary field. 
\nomenclature{$f^*$}{  denotes an auxiliary field  }%

In summary, the method is an optimization of the previous implementation \cite{miller_campylotic}, where the LBM velocity lattice vectors are reduced from $D3Q41$, having third and fifth nearest neighbors, to $D3Q19$. 2$^{nd}$-order convergence to the NSE is achieved  by changing the lattice Boltzmann equations. Essentially, adding supplementary contributions to the forcing term Eq.~(\ref{eq:forcing-C}) and thus the streaming step, which automatically cancel out spurious terms at relevant order in the NSE. Historically there has been other approaches to this, such as redefining the distribution function to regularize the lattice as proposed by Latt et al. \cite{latt_2006}, our scheme includes a change at the level of the streaming step.

As a consequence of this correction, the natural contribution of the $D3Q19$ lattice cannot be removed completely to all orders, so in a general coordinate system, higher order moments might not be isotropic and lead to some spurious effects. However, the covariant NSE are recovered correctly  to the same order as before  (see Appendix~\ref{app:chapmanenskog}). The advantages of the 2$^{nd}$-order method are computational efficiency and much simplified boundary implementations, important for the FSI coupling.

\section{FEM shell}
\subsection{Continuum description of thin elastic objects}
In this study, the FEM was used to simulate thin elastic objects. Specifically, subdivision shells developed by Cirak et al \cite{cirak1,cirak2} based on the Kirchhoff-Love theory were implemented due to their ability to resolve geometries undergoing large deformations with complex boundary conditions \cite{shellcode_thinsheets,shellcode}. These shells rely on subdivision surfaces to generate a smooth well-defined \textit{limit surface} (LS) from a coarse finite-element triangulation of the \textit{control mesh} (CM), instead of the aggregation of non-conforming local patches derived from traditional methods. This limit surface is constructed using Loop's recursive refinement \cite{loop} scheme generalized by Stam's eigendecomposition \cite{stam}. The basis functions used in this methodology are 12 quartic splines that produce the exact interpolated infinitely refined limit surfaces that are globally $C^2$ continuous except at a small number of irregular vertices where they are $C^1$. This allows for homogenous interfaces between elements due to the consistency between the geometrical and finite-element representation. More importantly, in the context of FSI, the fluid solver requires $C^2$ continuity whenever $\Gamma^i_{jk}$ is calculated. Furthermore, this class of shell elements has already been extended to include anisotropic growth \cite{roman_nature} and orthotropic behavior, which are necessary for simulating elastic objects in a biological context. 

\begin{figure}
\includegraphics[width=3.7in, height=1.7in]{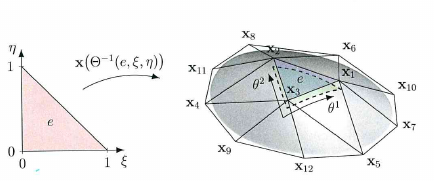}
\caption{\label{fig:limiting_surface} \textbf{Control mesh and subdivision surface.} The nodal coordinates map from the standard triangle (left) to the control mesh(solid lines) to the limit surface (dotted lines) of the triangular domain is obtained as a linear superposition of the 12 shape functions $N$ with the corresponding vertices $x_I$ as weights Eq.~(\ref{eq:vertices}). (extracted from \cite{roman_thesis})}  
\end{figure}

The fundamental difference in this case compared to traditional finite elements, is that the middle surface locally approximates the mesh nodal positions $x$ rather than interpolating them as seen in Fig.~\ref{fig:limiting_surface} as 
	\begin{equation}	\label{eq:vertices}
	x (\theta^1,\theta^2)=\sum^{12}_{I=1} x_I N_I(\xi, \eta),   
	\end{equation}
where $\theta^1$ and $\theta^2$ are the curvilinear coordinates of the shell element, $N_I$  the basis functions and $\xi \>= 0$ and $\eta \>= 0$ are the natural coordinates of the standard triangle.
\subsection{Thin Shell Mechanics}
The Kirchhoff-Love theory for thin shells uses the common stress-resultant formulation. In this formulation, the stresses are integrated analytically over the thickness of the shell, giving a resultant stress on its middle surface $\bar{\Omega}$. The shell is considered to be hyperelastic, and therefore uses the St. Venant-Kirchhoff constitutive material law. This allows for the use of the Koiter energy density functional; an effective elastic potential energy per unit surface area that links the kinematics and the energetics of the system  as
\begin{equation}
W = \frac{E H^{\alpha\beta\gamma\delta}}{1-\nu^2} \bigg[ h \alpha_{\alpha\beta} \alpha_{\gamma\delta} + \frac{h^3}{12} \beta_{\alpha\beta} \beta_{\gamma\delta} \bigg] \, ,
\end{equation}
where $E$ is the Young's modulus, $\nu$ is the Poisson's ratio, $H$ is the elasticity tensor, $h$ is the shell thickness, $\alpha$ is the membrane strain tensor, and $\beta$ is the bending strain tensor. The membrane and bending stress resultants are
\begin{equation}
\begin{split}
n^{\alpha\beta} &= \frac{\partial W}{\partial \alpha_{\alpha\beta}} = \frac{E H^{\alpha\beta\gamma\delta}}{1-\nu^2} h \alpha_{\gamma\delta} \, ,\\
m^{\alpha\beta} &= \frac{\partial W}{\partial \beta_{\alpha\beta}} = \frac{E H^{\alpha\beta\gamma\delta}}{1-\nu^2} \frac{h^3}{12} \beta_{\gamma\delta} \, .
\end{split}
\end{equation}
The total mechanical energy $\Pi$ of the Kirchhoff-Love \cite{Love} shell with total Lagrangian displacement of the middle surface $u= x-\bar{x}$, where $\bar{x}$ is the undeformed nodal position of the middle surface, with applied loads $q$ per unit surface area is therefore
\begin{equation}
\Pi [u] = \int_{\bar{\Omega}} h \r \dot{u} \cdot \dot{u} d \bar{\Omega} + \int_{\bar{\Omega}} W[u]\bar{\Omega} - \int_{\bar{\Omega}} q \cdot u \bar{\Omega} 
\end{equation}  
where $\dot{u}= \del u / \del t$ is the velocity field. $\Pi$ is then formulated as a variational statement, and solved by approximation as a discrete minimization problem. Therefore, in broad terms, the method solves for an elastic, kinetic and pressure energy Lagrangian. The time integration is performed using explicit dynamics with the \textit{constant-average acceleration method}. 
\section{Fluid Structure interaction \label{sec:FSI}}


\tikzset{
desicion/.style={
    circle,
    draw,
    text width=4em,
    text badly centered,
    inner sep=0pt
},
block/.style={
    thick,
    rectangle,
    draw=black!300,
    text width=23em,
    text centered,
    rounded corners
},
blocklb/.style={
    rectangle,
    dotted,
    draw=black!300,
    text width=23em,
    text centered,
    rounded corners
},
blockcouple/.style={
    rectangle,
    dashed,
    draw=black!300,
    text width=23em,
    text centered,
    rounded corners
},
blockshell/.style={
    rectangle,
    very thick,
    draw=gray!1000,
    text width=23em,
    text centered,
    rounded corners
},
connector/.style={
    -latex
},
rectangle connector/.default=-2cm,
straight connector/.style={
    connector,
    to path=--(\tikztotarget) \tikztonodes
}
}

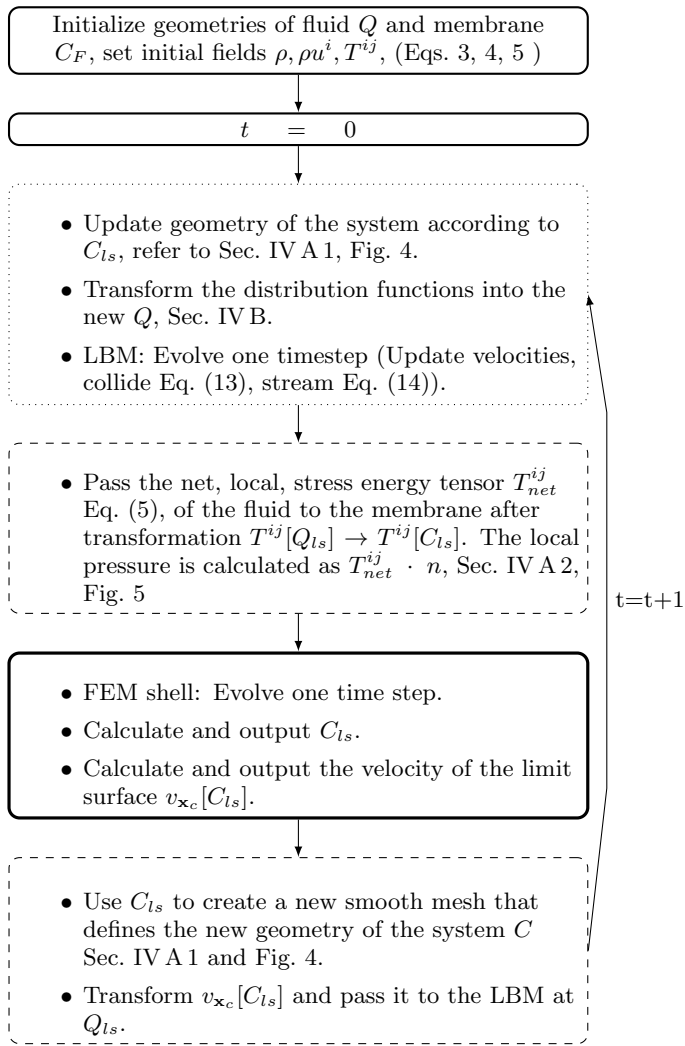
\begin {figure}
\begin{tikzpicture}
\matrix (m)[matrix of nodes, column  sep=2cm,row  sep=5mm, align=center, nodes={rectangle,draw, anchor=center} ]{
    |[block]| {Initialize geometries of fluid $Q$ and membrane $C_F$, set initial fields	$\rho, \rho u^i, T^{ij}$, (Eqs.~\ref{eq:fields1},~\ref{eq:fields2},~\ref{eq:fields3} )	
     }          
    & 
    \\
    |[block]| {$t=0$}          
    &    
     \\
   |[blocklb]| {
   \begin{itemize}
   \item Update geometry of the system according to $C_{ls}$, refer to Sec.~\ref{sec:interpolation}, Fig.~\ref{fig:interpolations}.
                \item Transform the distribution functions into the new $Q$,  Sec.~\ref{sec:transform_f}.
    		    \item LBM: Evolve one timestep (Update velocities, collide Eq.~(\ref{eq:collision}), stream Eq.~(\ref{eq:streaming})).
   \end{itemize} 		   
   }    
   &  
    \\
    |[blockcouple]| 
    {    \begin{itemize}

	 \item Pass the net, local, stress energy tensor $T^{ij}_{net}$ Eq.~(\ref{eq:fields3}), of the fluid to the membrane after transformation $T^{ij}[Q_{ls}] \rightarrow T^{ij}[C_{ls}]$.  The local pressure is calculated as $T^{ij}_{net}\cdot n$,  Sec.~\ref{sec:force_vel_coupling}, Fig.~\ref{fig:force_vel_coupling}
       \end{itemize} 		   
 
      }     
     &    
 \\
          |[blockshell]| {
 \begin{itemize} 
 \item FEM shell: Evolve one time step.  
 \item Calculate and output $C_{ls}$.
 \item Calculate and output the velocity of the limit surface $v_{\textbf{x}_c}[C_{ls}]$.
             \end{itemize}

      } 
&\\  
    |[blockcouple]| 
    
    { \begin{itemize}    
    \item Use  $C_{ls}$ to create a new smooth mesh that defines the new geometry of the system  $C$ Sec.~\ref{sec:interpolation} and Fig.~\ref{fig:interpolations}.  
    \item Transform $v_{\textbf{x}_c}[C_{ls}]$ and pass it to the LBM at $Q_{ls}$.    
            \end{itemize}
       	}  
&  \\       
};
\path [>=latex,->] (m-1-1) edge (m-2-1);
\path [>=latex,->] (m-2-1) edge (m-3-1);
\path [>=latex,->] (m-3-1) edge (m-4-1);
\path [>=latex,->] (m-4-1) edge (m-5-1);
\path [>=latex,->] (m-5-1) edge (m-6-1);
\draw [>=latex,->] (m-6-1.east) --(3.1,-3.5) -- node [right] {t=t+1} (3.1,1.5) -- (m-3-1.east) ;
\end{tikzpicture}

\caption{\label{fig:flowchart} \textbf{Method summary flow chart.} Thin solid line$\Rightarrow$ initialization, dotted line$\Rightarrow$ LBM, dashed line $\Rightarrow$ FEM shell, thick solid line $\Rightarrow$ coupling.}
\end{figure}


\begin {figure}
\includegraphics[width=1.01\columnwidth,trim={0 2.2cm 0 0}, clip]{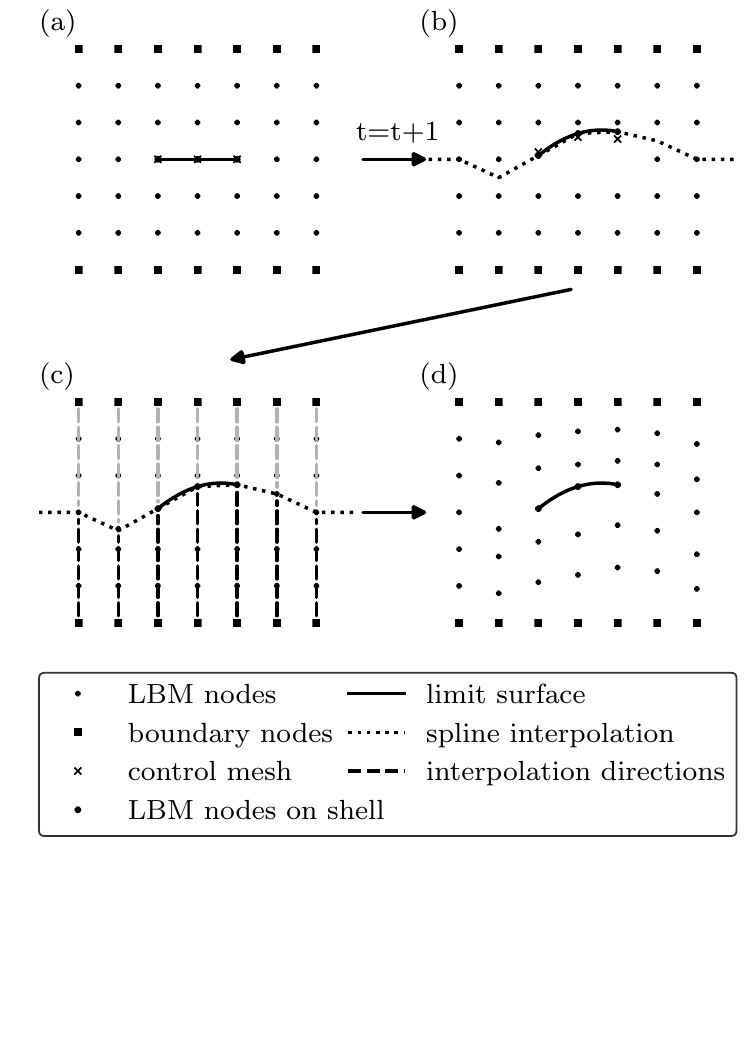}

\caption{\textbf{Position coupling diagram.} \textbf{(a)} Internal moving boundary.  \textbf{(b)} Cubic spline solution used to extend the shape of the  FEM shell for continuity, note that  $Q_{ls}=C_{ls} \neq C_{cm}$. \textbf{(c)} Interpolation carried out along orthogonal directions to the FEM shell, cubic spline and linear in case of interior and exterior moving boundaries respectively. \textbf{(d)} Final mesh configuration.}
\label{fig:interpolations} 
\end{figure}

Prior to the description of the coupling of the LBM and FEM domain, consistent notations are summarized:
\begin{itemize}

\item  The Cartesian basis vectors are $\textbf{x}_c=(x_c, y_c, z_c)$, which describe the Cartesian parametrization space $C(\textbf{x}_c)$, used in the FEM domain at time $t$. The subspace of the Cartesian lattice points on the FEM shell and the limit surface are denoted by $C_F(\textbf{x}_c)$ and $C_{ls}(\textbf{x}_c)$ respectively. The corresponding velocity vectors are expressed  by $v_{\textbf{x}_c}$.
\item The basis vectors $\textbf{x}=(x,y,z)$, describe the curved space manifold $Q(\textbf{x})$, used in the LBM domain. The subspace of the curved space manifold lattice points of the LBM on the limit surface is $Q_{ls}(\textbf{x})$ which is equivalent to $C_{ls}(\textbf{x}_c))$. The corresponding velocity vectors are expressed by $u_{\textbf{x}} \equiv u^i$. 

\item Coordinate transformations would be denoted by $\rightarrow$ and a quantity $A[M]$ is the transformed value of A on specific manifold $M$.
\end{itemize} 

The complete simulation procedure is summarized in Fig.~\ref{fig:flowchart}.
 \nomenclature{$t$}{time step }%
 \nomenclature{$x_c, y_c, z_c$}{ Cartesian basis vectors}%
 \nomenclature{$x,y,z$}{ LBM basis vectors in Sec.~I-V.A or simulation dimensions in Sec.~V.B-VI   }%
 \nomenclature{$C$}{ Cartesian parametrization space}%
 \nomenclature{$Q$}{  LBM curved space manifold }%
 \nomenclature{$C_F$}{Cartesian lattice points on FEM shell }%
 \nomenclature{$C_{ls}$}{ Cartesian lattice points on limit surface }%
 \nomenclature{$Q_{ls}$}{Curved space lattice points of the LBM on the limit surface }%
 \nomenclature{$A[M]$}{  transformed value of A on manifold M }%
 \nomenclature{$LS, CM$}{  Limit surface and control mesh of FEM shell }%

\subsection{ Coupling of the LBM and the FEM shell  }
  The interaction of the FEM shell to the fluid is achieved by direct coupling of the local positions, velocities and forces of the two methods. To be coupled, all the variables are coordinately transformed between the Cartesian and the LBM curved space coordinates.  In addition, due to the dynamic geometry the distribution functions of the LBM need to be transformed after each time-step, this is explained in Sec.~\ref{sec:transform_f}. 
\subsubsection{\label{sec:interpolation} Position coupling }
 
The shell position always resembles moving boundaries to the fluid solver. Unfortunately, the coordinate map $\vec h$ has to be updated respectively. Not only the entire domain needs to be covered but also connectivity needs to be conserved. Furthermore, for numerical accuracy, the chart needs to be at least $C^2$ differentiable with moderate gradients. This ensures a smooth metric and Christoffel symbols variation, which appear in the forcing terms of the LBM, Eqs.~(\ref{eq:forcing-A},~\ref{eq:forcing-B},~\ref{eq:forcing-C}). Note that not one unique interpolation method in-between spatial limits can be universally applied: 
\begin{itemize}
\item[(i)] In the case of an exterior FEM shell boundary with small deformations, linear interpolation is sufficient and can be implemented for a chart  $\vec h = \textbf{x}$ relative to a Cartesian grid  $\textbf{x}_c$ (domain $0 \leq \textbf{x} \leq L_i$) as, $
\textbf{x} (\textbf{x}_c) = \textbf{x} (0)+ \textbf{x}_c(\textbf{x} (L_i)-\textbf{x} (0) )/L_i$, where $i$ spans the 3-dimensional space as before.
\nomenclature{$\vec h $}{ Chart, physical positions of nodes }%
\item[(ii)] For an interior boundary a cubic spline interpolation is carried out along orthogonal directions to the FEM shell as in Fig.~\ref{fig:interpolations}(c) and (d). In addition it might be necessary to fit a cubic spline solution along the cross-section that overlaps with the FEM shell, which is then used to update the positions of that cross-section, Fig.~\ref{fig:interpolations}(b) and (c). This ensures $\mathcal{C}^2$ differentiability between the lattice nodes. 
\end{itemize}
For both cases, the LBM lattice nodes $Q_{ls}$ at each time step, are positioned on the LS. Once $\vec h$ is created, it is passed to the LBM as a geometry update, and all distribution functions are transformed to the new curved space manifold, $\tilde{Q}$ (see Sec.~\ref{sec:transform_f}). 
\subsubsection{\label{sec:force_vel_coupling} Velocity and force coupling }
\begin {figure}
\includegraphics[width=1.0\columnwidth,trim={6cm 6.5cm 5cm 1.5cm}, clip]{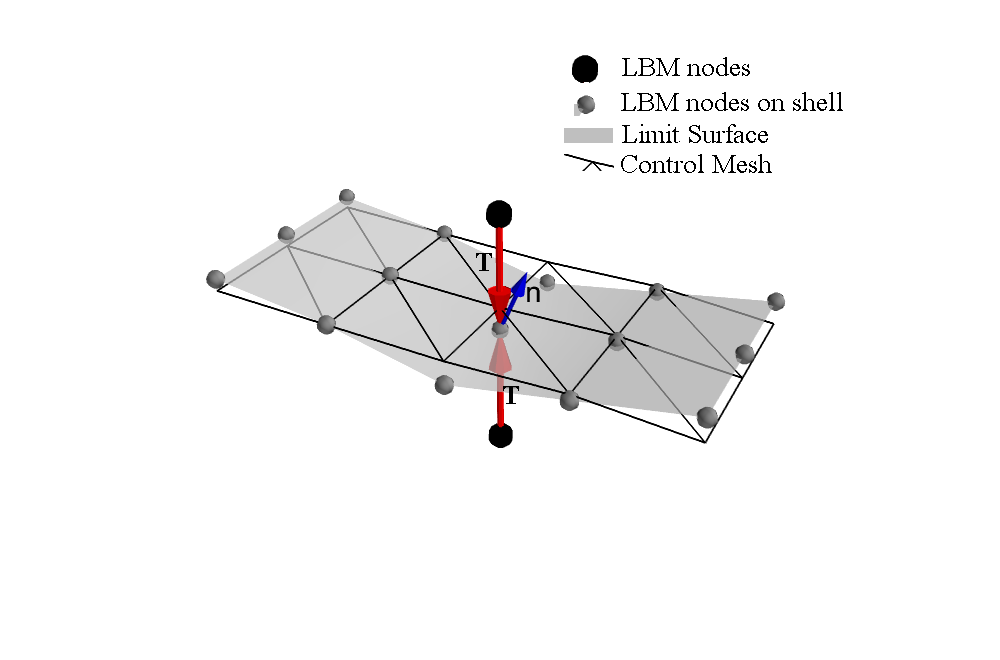}
	\caption{\textbf{Velocity and Force coupling.} The transformed velocities of the LS, $v_{\textbf{x}_c}[C_{ls}] \rightarrow u_\textbf{x}$ and then directly using $u_\textbf{x}$ as the new LBM velocity on the  $Q_{ls}$ nodes. The net stress-energy tensor $T^{ij}$, at each point on the LS, $T^{ij}[C_{ls}]$ is calculated from the LBM nodes directly above and below. The pressure is calculated as $T^{ij}_{net} \cdot n$.}
\label{fig:force_vel_coupling}
\end{figure}
Referring to Fig.~\ref{fig:force_vel_coupling}, the velocity of the LS, $v_{\textbf{x}_c}[C_{ls}]$  is coupled by $v_{\textbf{x}_c}[C_{ls}] \rightarrow u_\textbf{x}$ and then using $u_\textbf{x}$ as the new LBM velocity on the  $Q_{ls}$ nodes; i.e we directly copy the transformed velocity of the shell to the fluid on the local nodes. The force of the fluid is coupled by calculating the net local $T^{ij}$ from both sides of the FEM shell, as well as the normal vector $n$ of the LS at the node. The fluid pressure $T^{ij}_{net} \cdot n$ is then added as an external pressure acting on the CM.  

\subsection{\label{sec:transform_f} Transformation of the distribution function and tensors under coordinate change}

\begin{figure}[ht]
\includegraphics[width=1.0\columnwidth]{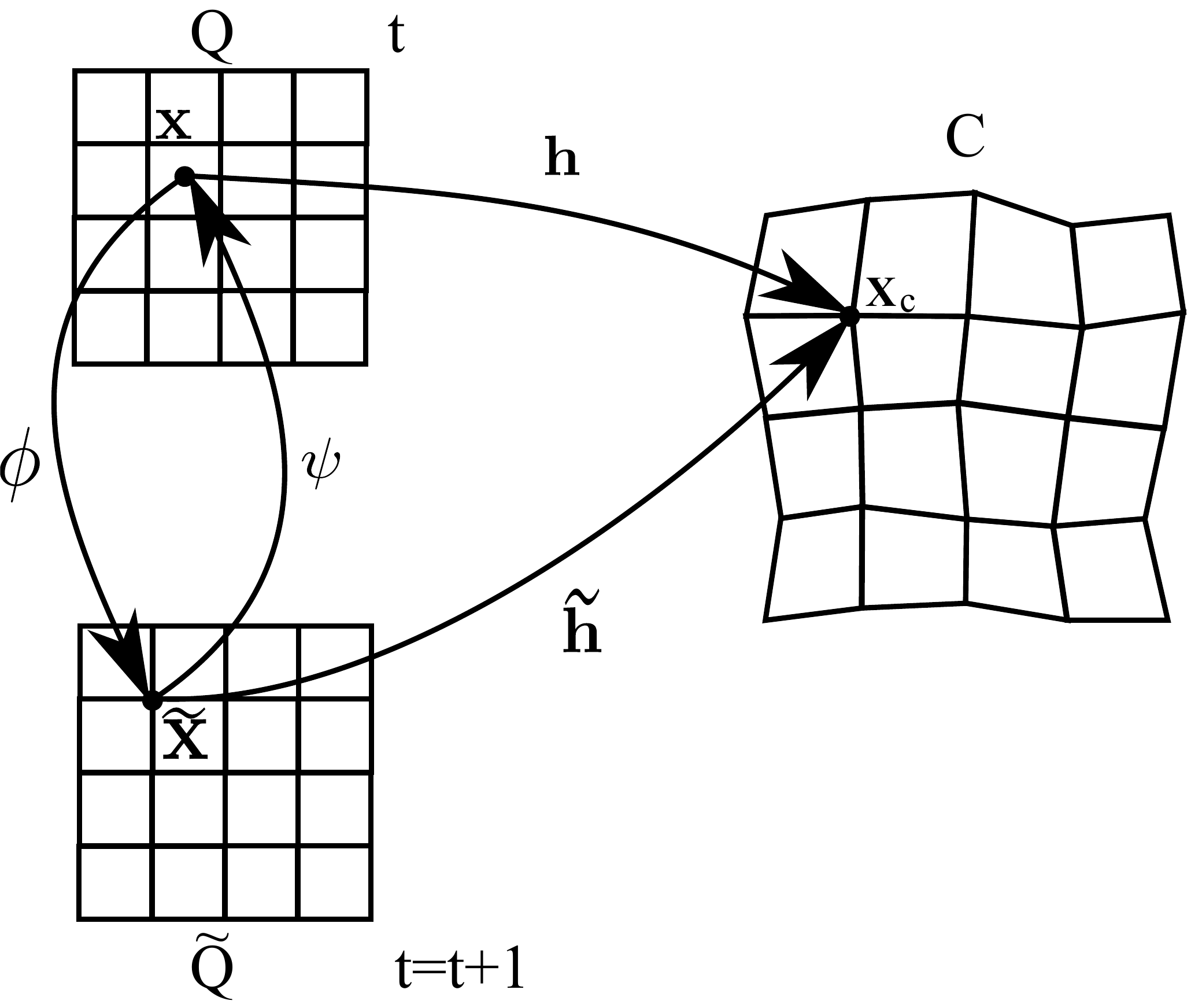}
\caption{\textbf{Chart transformation between two charts.} $\mathbf h : Q \rightarrow C$ and  $\widetilde{\mathbf h} : \widetilde Q \rightarrow C$ mapping from different parametrization spaces $Q$ and $\widetilde Q$ to the manifold $C$.}
\label{fig:chart_transformation}
\end{figure}

There are two types of coordinate transformation of interest to this work. The first corresponds to the transformation of variables between the LBM and Cartesian manifolds for coupling purposes as introduced in Sec.~ \ref{sec:FSI} or as shown on Fig.~\ref{fig:chart_transformation}, i.e. $Q \rightarrow C$ and $C \rightarrow Q$. The second concerns purely the LBM after a geometry change. More specifically the distribution functions need to be transformed to the new LBM manifold, i.e.  $\tilde{Q}\rightarrow Q$.

Let $\mathbf h : Q \rightarrow C, \mathbf h : \mathbf x \mapsto \mathbf{x}_c= \mathbf h(\mathbf x)$ be the chart that maps the rectangular grid $Q$ to the Cartesian manifold $C$ (e.g. at time $t$).
Let $\widetilde{\mathbf h} : \widetilde Q \rightarrow C, \widetilde{\mathbf h} : \widetilde{\mathbf x} \mapsto \mathbf{x}_c = \widetilde{\mathbf h}(\widetilde{\mathbf x})$ be a different chart (e.g. at time $(t+1)$). Then the tensor field $T^{i_1, i_2, ..., i_n}$ can be expressed by the Cartesian field $T^{i_1, i_2, ..., i_n}$ as:
\begin{align}
\label{eq:vector_transform}
T^{i_1, ..., i_n}(\mathbf x) = \pdiff{h^{i_1}}{x_c^{k_1}} \cdots \pdiff{ h^{i_n}}{x_c^{k_n}} T^{k_1, ..., k_n}(\mathbf{x}_c).
\end{align}
E.g. for a velocity field   $v^i(\mathbf x) =(\del h^i/\del x_c^a) v^a(\mathbf{x}_c)$.

The transition maps which relate the points in $Q$ to the corresponding points in $\widetilde Q$ (and vice versa) are given by:
\begin{align*}
	\boldsymbol\phi : Q \rightarrow \widetilde Q, 
	\qquad \boldsymbol\phi : \mathbf x \mapsto \widetilde{\mathbf x} = \widetilde{\mathbf h}^{-1}(\mathbf h(\mathbf x)), \\
	\boldsymbol\psi : \widetilde Q \rightarrow Q, 
	\qquad \boldsymbol\psi :  \widetilde{\mathbf x} \mapsto \mathbf x = \mathbf h^{-1}(\widetilde{\mathbf h}(\widetilde{\mathbf x})).
\end{align*}
 \nomenclature{$\psi, \phi $}{ Transition maps }%
(see Fig.~\ref{fig:chart_transformation}).
With these transition maps, the tensor fields $\widetilde T^{i_1, ..., i_n}$ defined in the new chart $\widetilde{\mathbf h}$ can be expressed by the `old' tensor fields $T^{i_1, i_2, ..., i_n}$ defined in the chart $\mathbf h$:
\begin{align*}
	\widetilde T^{i_1, ..., i_n}(\widetilde{\mathbf x}) = \pdiff{\phi^{i_1}}{x^{k_1}} \cdots \pdiff{\phi^{i_n}}{x^{k_n}} T^{k_1, ..., k_n}(\mathbf x),
\end{align*}
where $\mathbf x = \boldsymbol\psi(\widetilde{\mathbf x})$. In order to carry out the transformation of the lattice Boltzmann distribution function $f_\l(x^i, t)$,  must be expressed in terms of some Tensor field which depends only on space. To this end, a Hermite polynomial expansion is carried out for $f_\l(x, t)$  similar to $f^{eq}$ in Eq.~(\ref{eq:expansion-equilibrium-distribution}), with expansion coefficients given by
\begin{align*}
	a_{(n)}^{i_1,...,i_n} = \sum_\l f_\l \HH_{(n)}^{i_1,...,i_n} \sqrt g.
\end{align*}
Written in this way, $f_\l$ depends on space only implicitly through its moments $a_{(n)} = a_{(n)}(x^i)$. Thus, it is sufficient to transform only the moments:
\begin{align*}
	\widetilde a_{(0)}(\widetilde{\mathbf x}) &= a_{(0)}(\mathbf x) \\
	\widetilde a_{(1)}^{i}(\widetilde{\mathbf x}) &= 
	\pdiff{\phi^{i}}{x^{l}}(\mathbf x)  \ 
	a_{(1)}^{l}(\mathbf x) \\
	\widetilde a_{(2)}^{ij}(\widetilde{\mathbf x}) &= 
	\pdiff{\phi^{i}}{x^{l}}(\mathbf x) \ 
	\pdiff{\phi^{j}}{x^{m}}(\mathbf x) \ 
	a_{(2)}^{lm}(\mathbf x). \\
\end{align*}
Since $\boldsymbol\psi(\widetilde{\mathbf x})$ will be off-grid in $Q$ in general, the values of the transformed moments have to be interpolated from the neighboring grid points. Finally, the transformed distribution function can be recomposed from the transformed moments:
\begin{align}
	\widetilde f_\l &= \frac{w_\l}{\sqrt g} \bigg( \widetilde a_{(0)} \HH_{(0)}
	+ \frac{1}{c_s} \widetilde a_{(1)}^{i} \HH_{(1)}^i
	+ \frac{1}{2! c_s^2} \widetilde a_{(2)}^{ij} \HH_{(2)}^{ij}+... \bigg).
\end{align}

In our implementation we include terms up to second order because they are enough to reproduce hydrodynamics.

It is instructive to clarify that this method uses a standard lattice, which is unaffected by the dynamic geometry. Instead, the chart characterizes the metric tensor, which then defines the new curved space coordinates. As a consequence, it is  sufficient that the distribution functions are  coordinate transformed on the nodes without the need of interpolation.

\section{Accuracy and stability verification of the method}
\subsection{Lattice Boltzmann}
As a consequence of implementing a  2$^{nd}$-order variation of the curved-space LBM we carry out some basic tests for accuracy validation.  Referring to Fig.~\ref{fig:flowchart}, in this subsection the method is implemented up to and including the 3$^{rd}$ box. Firstly in order to verify the hydrodynamics, the standard Poiseuille flow is investigated with different resolutions. This is done by applying a constant external force  $F_{ext}$ across the domain and comparing the steady state solution (when $ \del(\r u^i )/ \del t = 0$ )  of the cross-sectional velocity profile to the analytical expression  $v_{x_c}= F_{ext} (x_c L_2^c-x_c^2)/(2 \nu)$.  $L_i^c$ and  $L_i$ denote the computational domain size in Cartesian and curved space coordinates respectively with $i=1,2,3$.

The next step is to introduce curvature to the system, i.e. some perturbation to the flat metric. An interesting example is the distortion of the transverse metric component $g_{yy}$ in flow direction $x$, which introduces shear components to the induced force. This is parametrized by $h'(y) =\epsilon \cos( \pi y / L_2 )$, where $\epsilon =0.1$. The complete metric is given by 
\begin{align*}
	g_{ij} = 
	\begin{pmatrix}
		1 & 0 & 0 \\
		0 & 1+\del^2_{y} h'(y)  & 0 \\
        0 & 0 & 1
	\end{pmatrix}.
\end{align*}

\begin{figure}
\includegraphics[width=1.0\columnwidth]{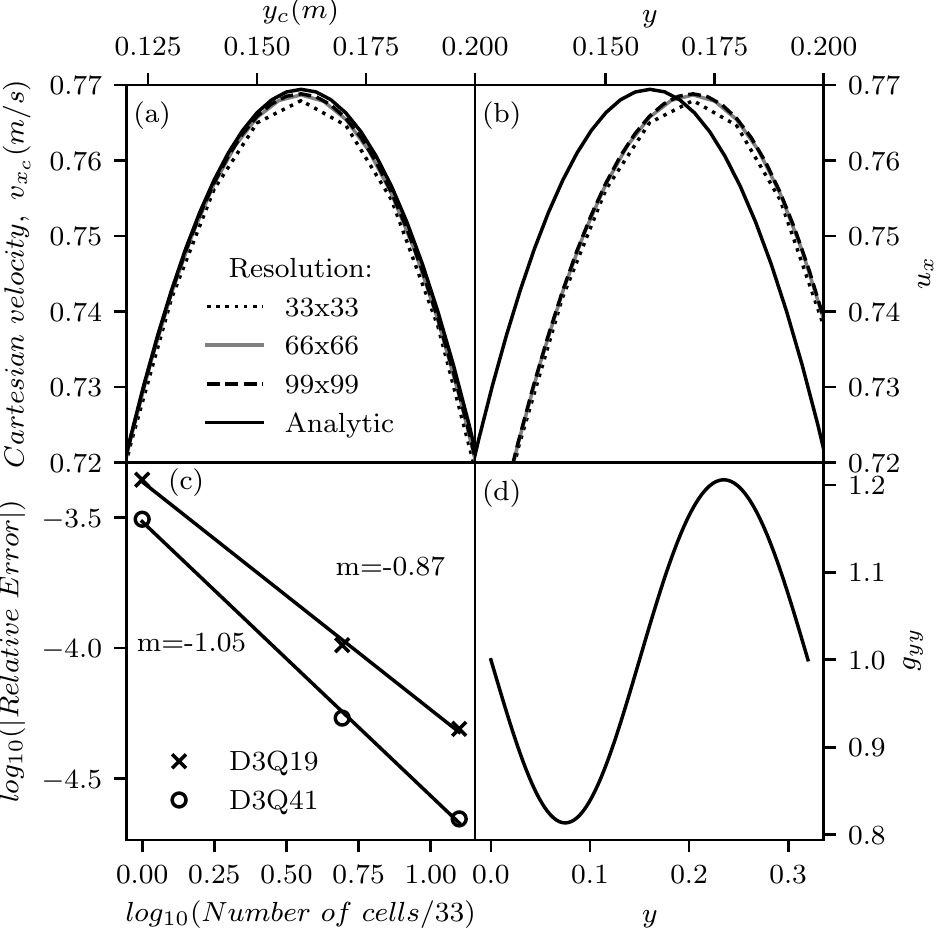}
\caption{ \textbf{Poiseuille velocity profile.}   \textbf{(a)} Analytic and numerical velocity profile in Cartesian coordinates ($v_{x_c}$) for a series of resolutions. \textbf{(b)} Velocity profile on the LBM curved space ($u_{x}$) compared to the Cartesian analytic solution. \textbf{(c)} Convergence of the relative error with increasing resolution for current $D3Q19$ and original $D3Q41$ LBM, $m$ denotes the gradient. \textbf{(d)} Deformed metric tensor component.}
\label{fig:z_test}
\end{figure}

The result is compared to the analytical solution both in  LBM and Cartesian coordinates, see  Fig.~\ref{fig:z_test}. The Cartesian velocity profile matches well the analytic solution as shown in Fig.~\ref{fig:z_test}(a).  The velocity profile when plotted in LBM coordinates appears significantly shifted relative to the analytic solution as it is not transformed (see  Fig.~\ref{fig:z_test}(b)). This is a strong indication that vector transformations are performed accurately.   The error for $v_{x_c}$ is shown to converge with increasing resolution. The equivalent result for the original $D3Q41$ LBM shows marginally faster convergence. This is expected as both methods recover NSE in the same order.

To further verify the accuracy of the method for anisotropic flow and dynamic deformations, the Taylor-Green vortex at Reynolds number $\approx$ 20 is implemented with a time-dependent metric. The velocities are initialized to $v_{x_c}(t_0)=v_0 \sin(2 \pi x_c/L_1^c) \cos(2 \pi x_c/L_1^c)$ and $v_{y_c}(t_0)=v_0 \cos(2 \pi y_c/L_2^c) \sin(2 \pi y_c/L_2^c)$ for some initial velocity amplitude $v_0$ and periodic boundary conditions. Defining $h'^i(\mathbf{x},t) =\epsilon \sin( \pi \mathbf{x} / L_i - \eta t), \epsilon =0.1 \ \eta=0.05$, the dynamic metric is realized as:

\begin{align*}
	g_{ij} = 
	\begin{pmatrix}
		1+\del^2_{x} h'^1(x, t) & 0 & 0 \\
		0 & 1+\del^2_{y} h'^2(y, t)  & 0 \\
        0 & 0 & 1
	\end{pmatrix}.
\end{align*}

\begin{figure}
\includegraphics[width=1.0\columnwidth]{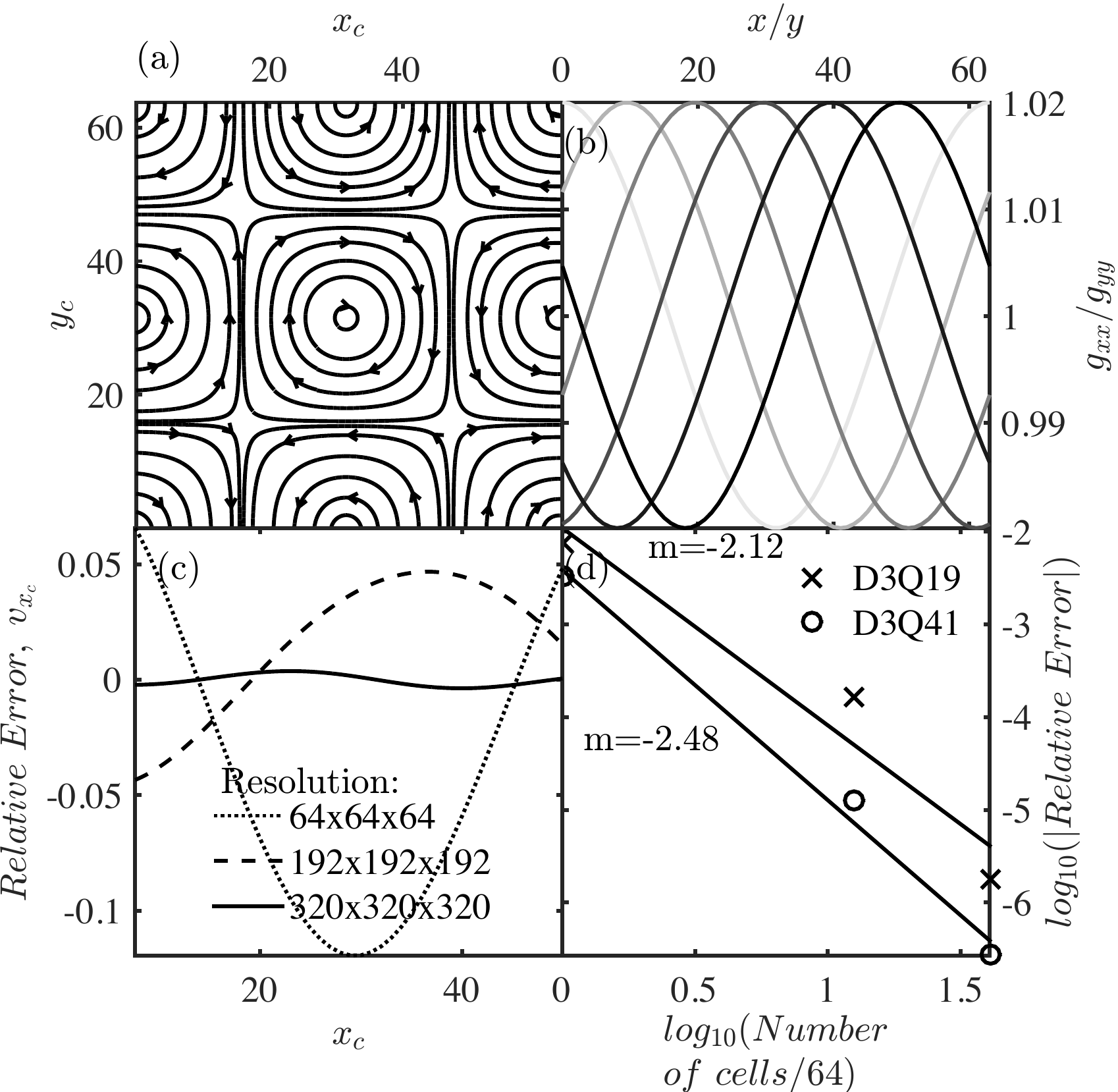}
\caption{ \textbf{Taylor-Green vortex.}  \textbf{(a)} Streamlines of the vortex in Cartesian coordinates ($u_{c}$) for time-step 50. \textbf{(b)} Deformed dynamic metric tensor components, time evolution in gray-scale. \textbf{(c)} Cartesian velocity component $u_{c1}$ relative error for a series of resolutions. \textbf{(d)} Convergence of the relative error with increasing resolution for current $D3Q19$ and original $D3Q41$ LBM, $m$ denotes the gradient.}
\label{fig:tgstudy}
\end{figure}

The analytic velocity in 2D can be calculated as  $v_{analytic}(\mathbf{x}_c,t)=v_{\mathbf{x}_c}(t_0) exp(-2 \nu t)$, where $\nu$ is the kinematic viscosity. The streamlines in Cartesian coordinates at a finite time-step are shown in  Fig.~\ref{fig:tgstudy}(a) and the dynamic metric in  Fig.~\ref{fig:tgstudy}(b). The relative error for $v_{x_c}$ is reduced with  increasing resolutions, see Fig.~\ref{fig:tgstudy}(c). The total average relative error for $v_{x_c}$  and $v_{y_c}$  are shown to converge with increasing resolution, see Fig.~\ref{fig:tgstudy}(d). Therefore, the ability of the method to work with dynamic and multi-component deformations is verified. The equivalent results for the original $D3Q41$ LBM demonstrate slightly better accuracy and convergence due to the higher isotropic moments when compared with the current method. 

Furthermore we carried out a series of calculation and transformation tests, namely the velocities are transformed between  $Q\rightarrow C\rightarrow Q$ using the vector transformation Eq.~(\ref{eq:vector_transform}), and the lattice derivatives calculation. The former directly affects the coupled quantities and the latter are fundamental in many calculations including Christoffel symbols, which are used in the forcing term of the LB equation. The velocity transformation has shown a perfect result up to the error of the derivative, which has shown quartic convergence with resolution increase (i.e. slope of Log(relative error)/Log(resolution refinement) $\approx -4$), see Appendix~\ref{app:derivative}.

\subsection{Coupled system}

\begin{figure}
\includegraphics[width=1.0\columnwidth]{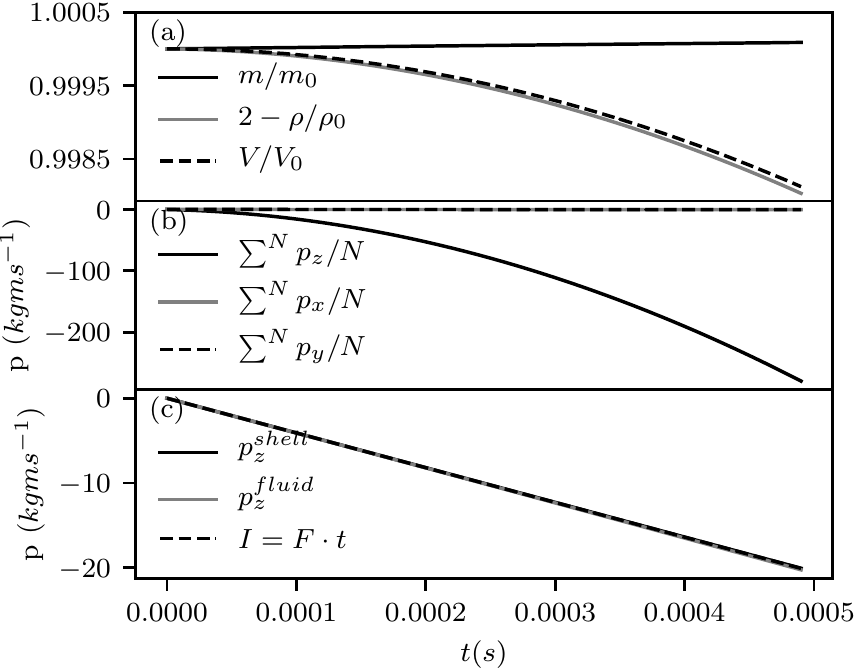}
\caption{\textbf{Mass conservation and momentum transfer}. \textbf{(a)} Normalized average mass, volume and density are plotted against time. Density increase is shown reversed for easy comparison with volume change. \textbf{(b)} The three independent momentum components are shown. \textbf{(c)} The total momentum (pressure) transfered to the membrane and fluid.}
\label{fig:piston_test}
\end{figure}

The accuracy of the method with respect to the basic conservation laws of mass and momentum is verified by a piston test with simple analytic solution. Henceforth, for clarity, we will refer to the simulation directions as $x, y ,z$ (Euclidean space in Cartesian basis, equivalent to $x_c,y_c,z_c$) and the domain size as $L_x, L_y, L_z$. Namely we simulate a rigid plate which is pressed down ($z$ direction) on the fluid. This is done for up to $0.2 \%$ volume change as the fluid is modeled as being very close to the incompressibility limit. The total mass of the fluid and momentum transfered are measured for each time step. As shown on the top graph of Fig.~\ref{fig:piston_test}
the normalized mass stays constant up to a $0.1 \%$ error. The conservation of mass comes about the increase in density as the volume decreases,  as shown in the top graph of Fig.~\ref{fig:piston_test}. Furthermore in the middle plot of Fig.~\ref{fig:piston_test}, the $x$ and $y$ average momenta are shown to be negligible compared to the $z$ component, as expected. Finally, the momentum change can be calculated analytically as $P \times t \times Area$, where $P$ denotes pressure. This analytic solution is compared to the transfer of momentum to the FEM shell and then from the shell to the fluid, which match up to a negligible error. 

\begin{figure}
\includegraphics[width=1.0\columnwidth]{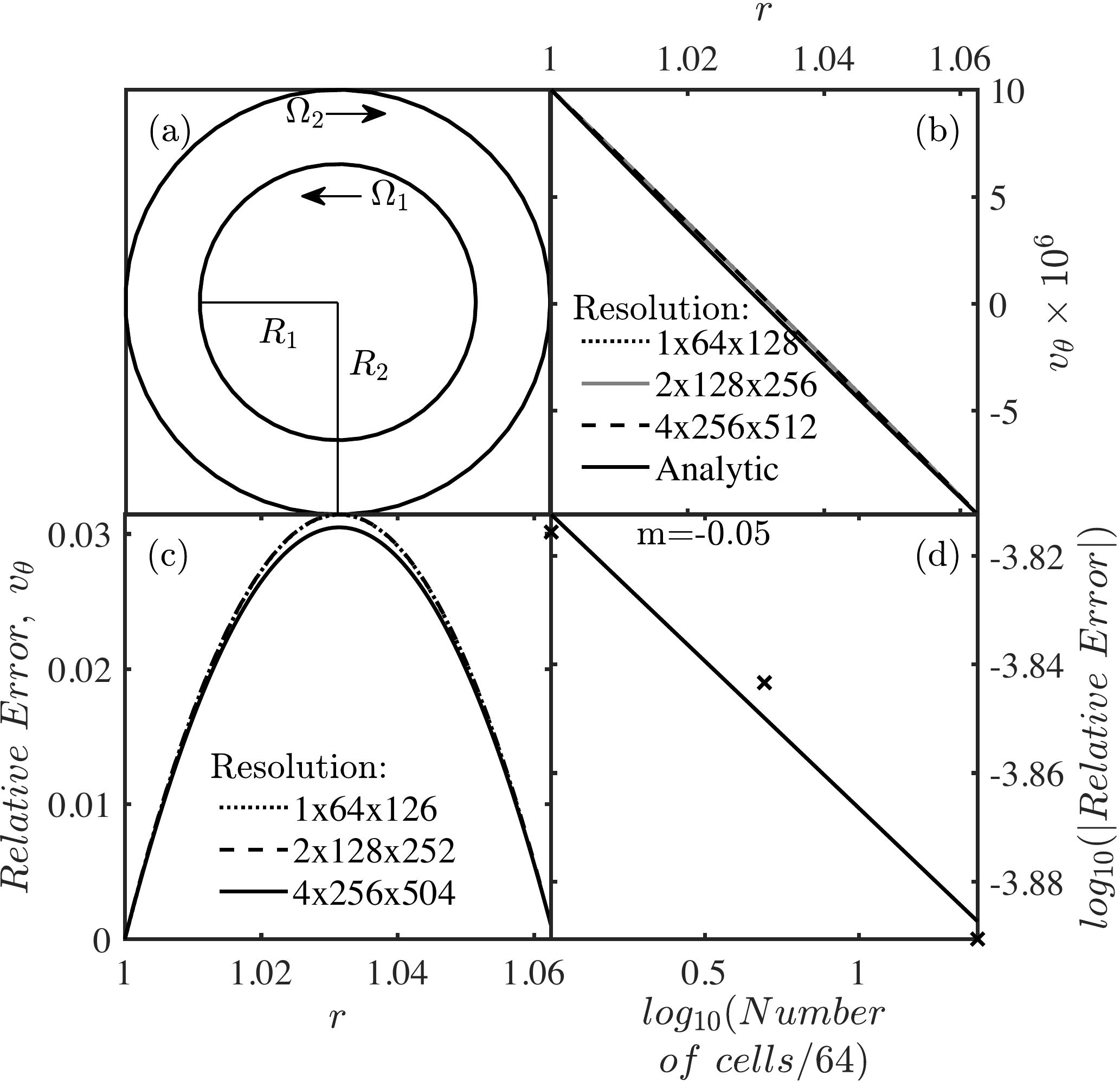}
\caption{  \textbf{Circular Couette flow.}  \textbf{(a)} Schematic diagram, laminar fluid flow between two oppositely rotating concentric cylinders. Symbols are defined in the main text. \textbf{(b)} Angular velocity profile. \textbf{(c)} Angular velocity relative error for a series of resolutions. \textbf{(d)} Convergence of the relative error with increasing resolution, $m$ denotes the gradient.}
\label{fig:ccflow}
\end{figure}

To further validate the accuracy of the method with respect to boundary slip the laminar circular Couette flow similar to the one in Ref.~\cite{stability} is simulated. The fluid velocity  depends on the tangential boundary motion, see Fig.~\ref{fig:ccflow}.  The analytic solution for the angular velocity $v_\theta$ is 
\begin{equation}
u_\theta(r)=C_1r+\frac{C_2}{r},
\end{equation}
where $C_1=(\Omega_2 R_2^2 - \Omega_1 R_1^2)/(R_2^2-R_1^2), \mathrm{ ~ and ~} C_2=(\Omega_2  - \Omega_1)R_1^2R_2^2/(R_2^2-R_1^2), ~ \Omega_1=1\times 10^{-6}, \Omega_2=-1\times 10^{-6}, R_1=1, R_2=1.063$ and they represent the inner and outer angular velocities, and the inner and outer radii respectively. The $v_\theta$ obtained at steady state for different resolutions is compared with the analytic expression as shown in Fig.~\ref{fig:ccflow}(b). The relative error for $v_\theta$ is reduced with increasing resolutions as shown in Figs.~\ref{fig:ccflow}(c)(d). The results indicate that the tangential velocity components of the coupled method are accurately implemented.

\begin{figure}
\includegraphics[width=1.0\columnwidth]{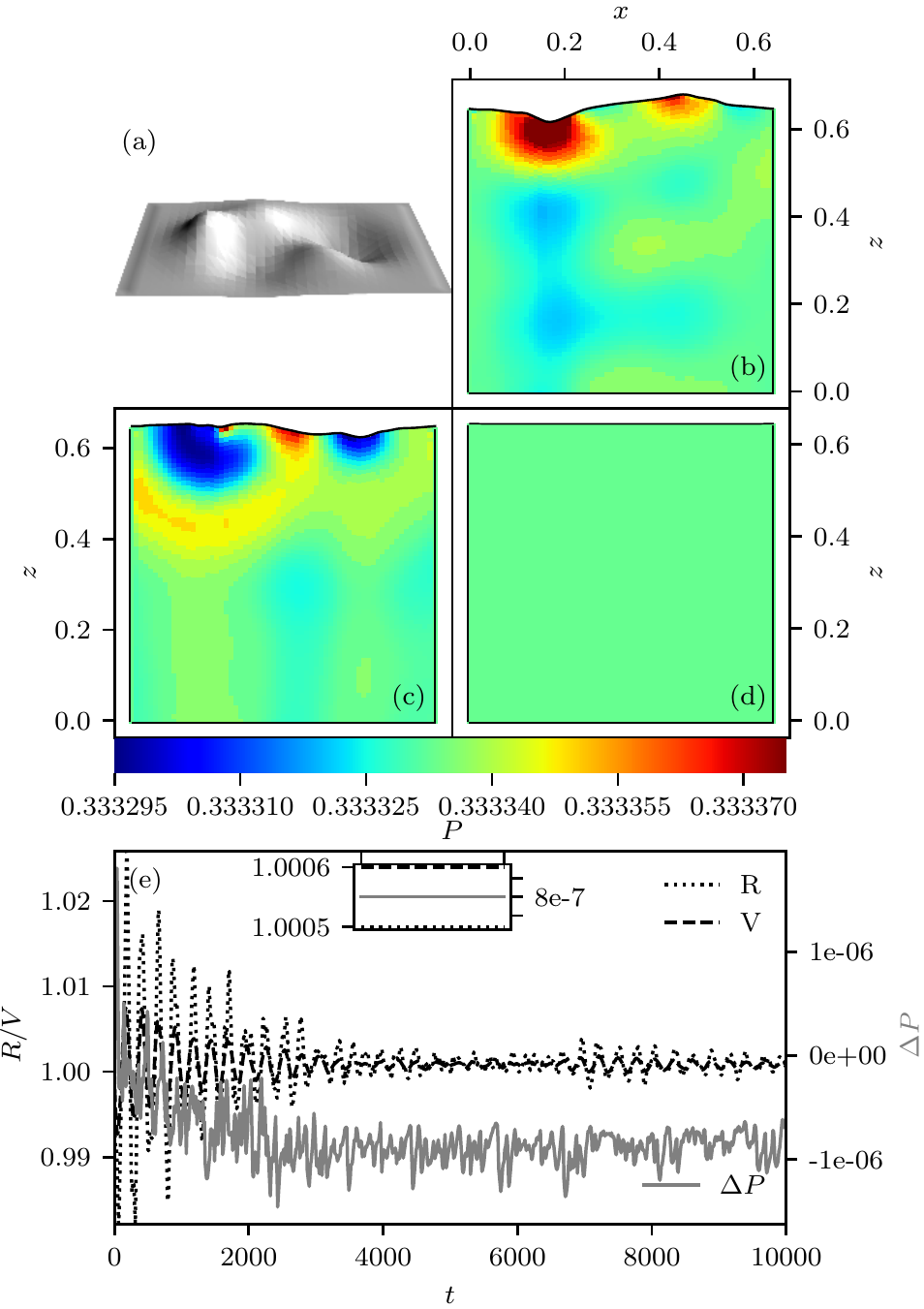}
\caption{  \textbf{Pressure maintenance and volume conservation.}  \textbf{(a)} Top surface FEM shell initial shape. \textbf{(b)(c)(d)} Cross-sectional pressure plot at time-step $200, ~ 500 ~ \mathrm{and} ~10^5$ respectively.  \textbf{(e)} Time-histories of the total normalized volume $V$, the  normalized FEM shell middle point location $R$ and relative pressure difference $\Delta P$, the inset shows the steady state values at time-step $10^5$.  }
\label{fig:bbtest}
\end{figure}

The pressure maintenance, and volume conservation accuracy are verified by a modified deformed balloon case as proposed in ~Refs.~\cite{stability2, stability3}. To this end, a closed rectangular 3-D system is simulated in contrast to the original 2-D circular membrane. A deformed FEM shell top surface and rigid walls are initialized with no-slip boundary conditions and uniform pressure. Fig.~\ref{fig:bbtest}(a) shows the initial shape of the deformed surface.  For this case the inside volume will be conserved and pressure may jump temporarily but equalize again. Due the equal pressures and fixed boundaries, the system is expected to reach a stationary steady state with a flat top surface. The pressure cross section at different time-intervals is shown in Figs.~\ref{fig:bbtest}(b)(c)(d).   The system oscillates until a stationary state is reached. The  normalized volume and pressure change time-histories are plotted in Fig.~\ref{fig:bbtest}(e), where the inset shows the steady state solutions at large $t$. The expected steady state volume and pressure are recovered to a high degree of accuracy.

\subsection{Stability analysis}

\begin{figure}
\includegraphics[width=1.0\columnwidth]{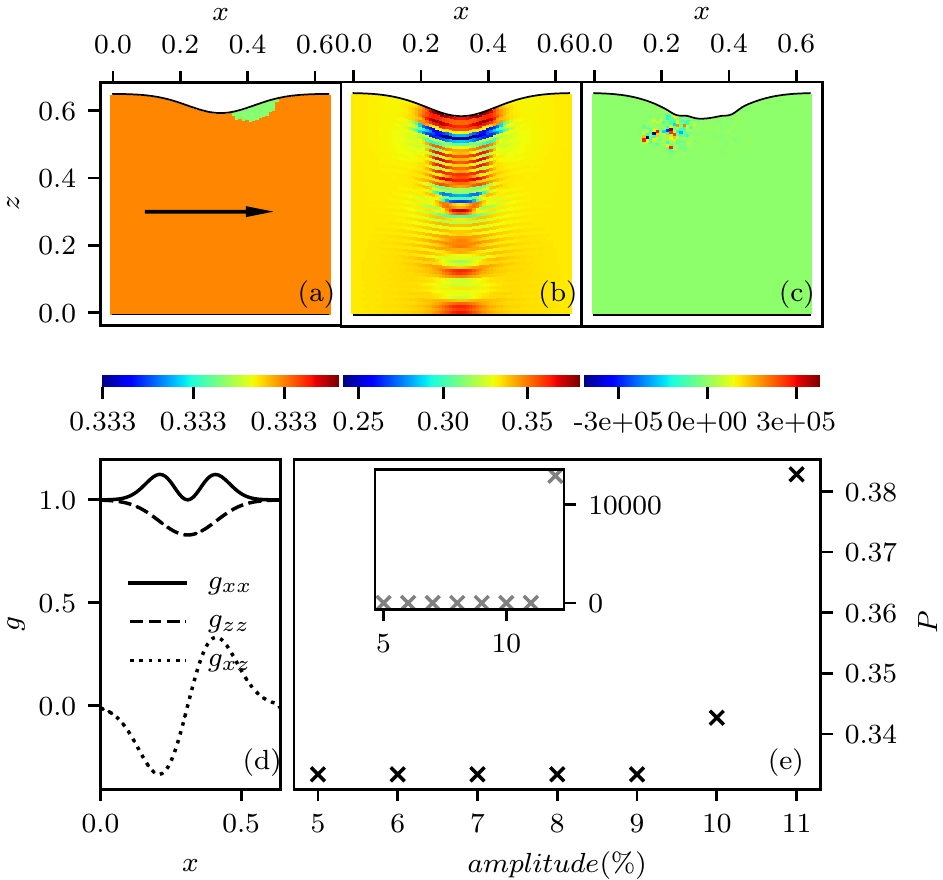}
\caption{  \textbf{Stability analysis}  \textbf{(a)} Pressure plot at steady state solution for amplitude of the concavity $9\%$ relative to $L_z$, the arrow indicates the flow direction.  \textbf{(b)} Pressure plot at steady state solution for amplitude  $11\%$. \textbf{(c)} Pressure plot at time-step 140 for amplitude  $12\%$. \textbf{(d)} Initial metric tensor components at the top boundary. \textbf{(e)} Maximum pressure at steady state for a range of amplitudes, the inset shows the maximum pressure at time-step 140 including amplitude $12 \%$.}
\label{fig:stability}
\end{figure}

 Large domain deformations appear to be the main cause of possible numerical instabilities of the method. Therefore, a stability analysis is carried out implementing a  rectangular channel  with a top deformable FEM shell, similar to the one in Sec.~\ref{sec:squarechannel}. The FEM shell is initialized with a Gaussian shaped  concavity as a heavy and stiff membrane to resist large changes, see Fig.~\ref{fig:stability}. The amplitude of the concavity is increased until no  steady state can be reached due to unmanageable numerical instabilities. A steady state solution, the appearance  of stable spurious effects and the emergence of instabilities in pressure are shown in Fig.~\ref{fig:stability}(a),~\ref{fig:stability}(b) and ~\ref{fig:stability}(c) respectively. In Fig.~\ref{fig:stability}(e) the maximum pressure within the domain at steady state is plotted against amplitude of the concavity. The simulation appears to be well behaved up to deformations of $10\%$ relative to the domain height, $L_z$. The metric, as plotted in Fig.~\ref{fig:stability}(e), indicates that the cross diagonal components deviate the most from the undeformed case, causing unstable forcing terms in Eqs.~(\ref{eq:CE-LB},\ref{eq:forcing-A},\ref{eq:forcing-B},\ref{eq:forcing-C}).

\section{Simulations}
Following the initial validation  the method is implemented in 2-D and 3-D scenarios. The first corresponds to a very common usage/validation of FSI methods, that is a moving flag in a non-laminar flow (Re$\approx 230$), which has been thoroughly studied in literature both experimentally  and numerically \cite{flag_exp}. The  flow induces an oscillatory motion to the flag, where the frequency of oscillations depends on the material, fluid, velocity and Reynolds number (Re). Additionally, this example allows for the exploration of the potential the method might have in non-laminar flows. The 3-D simulation, we implement is a simple square channel with an inlet outlet, 3 solid, non slip boundaries  and a deformable shell (no-slip) on top which is pressed down by an applied uniform pressure.
\subsection{\label{sec:flag} 2-D simulation, Flag on a pole}
 
\begin{figure}

\includegraphics[width=0.8\columnwidth]{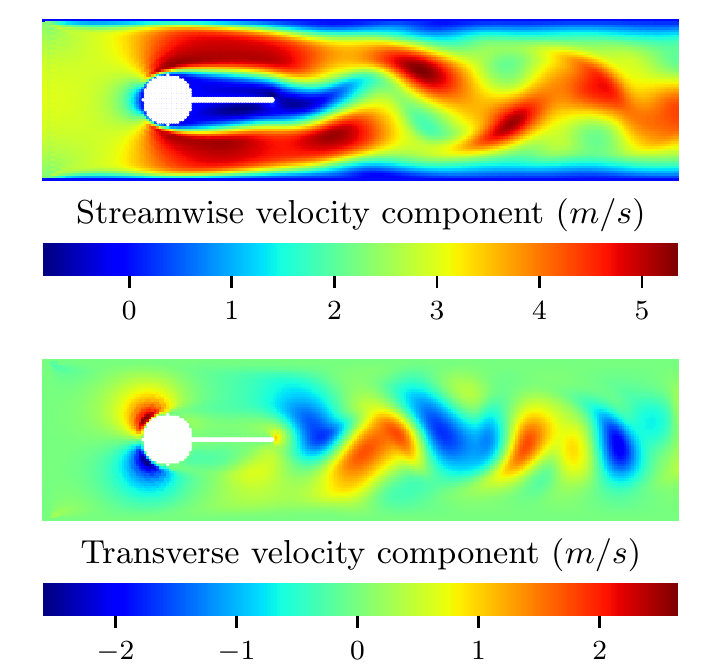}
\caption{ Flag and cylindrical pole are shown in white. Parabolic velocity inlet on the left and open outlet on the right. No-slip condition applied to the other boundaries, flag and pole. Snapshot taken during non-laminar flow(Re $\approx 230$).  }
\label{fig:flag_vel}
\end{figure}

A flag on a pole scenario is implemented as seen in Fig.~\ref{fig:flag_vel}. Specifically, a long air channel
(density $\rho_{air} =1\mathrm{kg/m}^3, \mathrm{~kinematic~viscosity}~\nu=0.00166\mathrm{m}^2/\mathrm{s},~L_x \times L_y = 257 \times 65 \mathrm{cm,~resolution~}  257 \times 65$) 
with a parabolic velocity profile at the inlet 
($v_{x}=4 \times \mathrm{m~s}^{-1}  (L_y^2-(y-L_y)^2) / L_y^2$) and
Re $\approx 230$ is simulated. Inlet and outlet are prescribed as open and for all other boundaries: top, bottom wall, flag and pole; no-slip condition is used. Open boundaries are implemented by extrapolation from the inner neighbors to the unknown distributions, where no-slip is achieved by fixing the velocity to zero on the boundary. The pole is a rigid cylinder with diameter 10cm, the flag has a thickness of 0.2cm, a length of 33cm and is made of EPDM-rubber (density: 1360$\mathrm{kg/m}^3$, Young's modulus: 16MPa, Poisson's ratio: 0.48). Additionally we impose some light damping. Interior boundary interpolation was implemented for this simulations as explained in Sec.~\ref{sec:interpolation}.

\begin{figure}
\includegraphics[width=1.0\columnwidth]{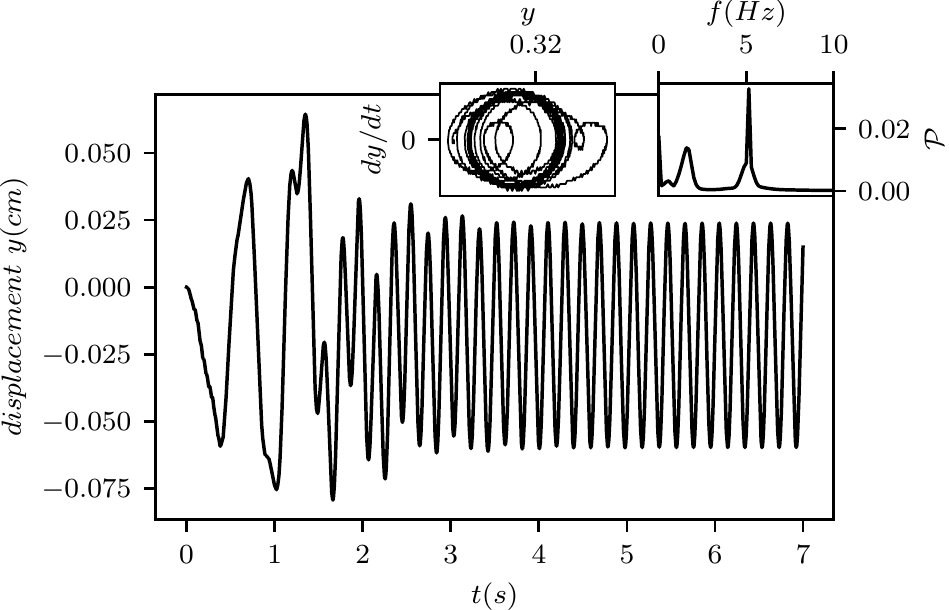}
\caption{\textbf{Flag tail oscillation.} Evolution of the cross-stream tail position, normalized by steady state amplitude. Initial irregular motion settles to regular periodic motion, after 4s. Left inset: phase plot of cross-stream displacement and velocity. Right inset: power spectrum of cross-stream displacement.}
\label{fig:flag_oscillations}
\end{figure}

The flow (Fig.~\ref{fig:flag_vel})  induces some instabilities to  the flag, which gradually settle to a periodic oscillation as in Fig.~\ref{fig:flag_oscillations}. The stable period of oscillation appears constant with a frequency $f \approx 5.1$Hz and an amplitude of $0.045cm$, as seen from the power spectrum $\mathcal{P}$ and displacement in Fig.~\ref{fig:flag_oscillations}. The second, lower frequency in the power spectrum is the dominant excitation of the initial instabilities that the flag experiences, which then decay.

These results can be compared to the FSI3 test as found in Ref.~\cite{Turek2006}, a proposal for FSI benchmarking by Turek et al.. In FSI3 the same dimensions and boundary conditions are used, Re $= 200$, and the shell to fluid density ratio is unity, which is much lower than in our case (i.e. 1360), for the reasons explained below. In FSI3, the steady state frequency $\approx 4.8Hz$ and the amplitude of oscillation $\approx 3.4cm$. The discrepancy in frequency is expected mainly due to the large difference in the shell to fluid density ratio and Reynolds number. The amplitudes are not comparable due to this large difference in the densities ratio.

The internal boundary implementation can be unstable to large and/or fast deformations for manageable resolutions, 
implying that in this case it does not converge to a steady state solution. Therefore, in order to  achieve a steady state solution, the current implementation is restricted to high shell to fluid density ratio. The instabilities appear to develop  in the 'expanded' half plane of the simulation domain (see Fig.~\ref{fig:interpolations}), due to not sufficiently smooth metric variations. The consequence of this is unstable forcing terms.

\begin{figure}
\includegraphics[width=1.0\columnwidth]{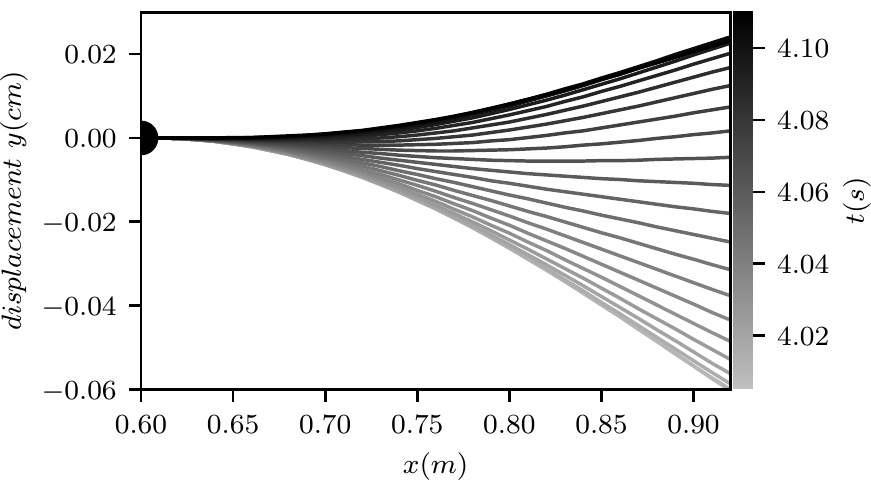}
\caption{\textbf{Flag Period.} The Flag's shape is plotted for one period of oscillation with resolution of 30 snapshots, time evolution in gray-scale. }
\label{fig:flag_period}
\end{figure}

Furthermore, as shown in Fig.~\ref{fig:flag_period}, the oscillation of the flag within one period, exhibits a motion around the clamped point. The oscillation shows similar qualitative behavior to the first simulation of \cite{Gilmanov2015} where a similar scenario with a high shell to fluid density ratio is implemented. The asymmetry in the oscillation comes about the slightly asymmetric velocity profile. This is also highlighted by the phase plot in Fig.~\ref{fig:flag_oscillations}.

\subsection{Square channel}
\label{sec:squarechannel}
\begin{figure}
\includegraphics[width=1.0\columnwidth,  trim={12cm 0cm 3cm 3cm },clip]{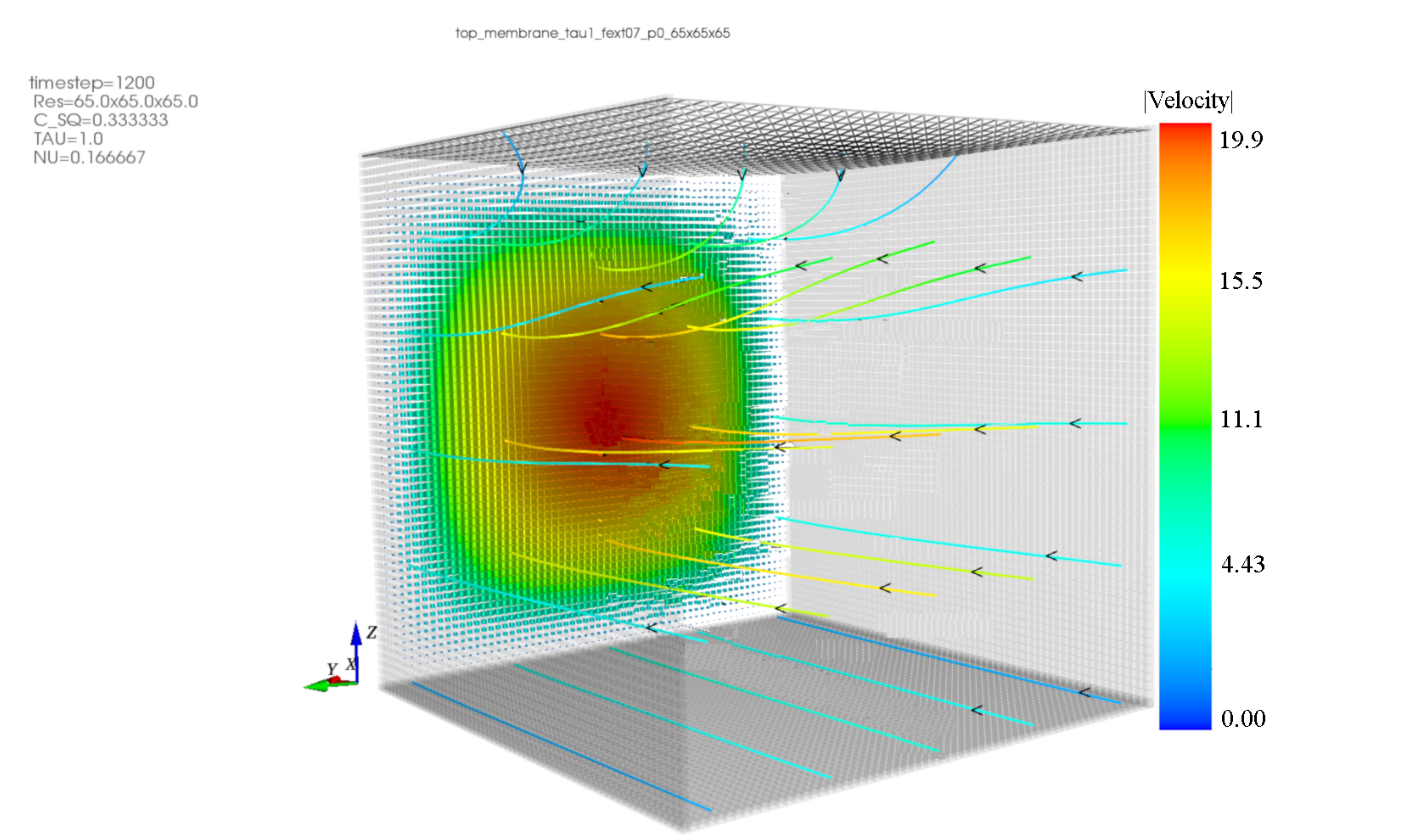}
\caption{\textbf{Square channel}. 3-D plot of velocity vectors and streamlines (Cartesian components). A pressure gradient is applied along the open, streamwise, $x$ boundaries. No-slip condition is applied to the other boundaries. Triangular mesh on top represents the deformable membrane, that is pushed down by a uniform pressure. Streamwise velocity exhibits a parabolic profile, which is affected by the membrane.}
\label{fig:TM}
\end{figure}

Moving to a truly 3-D implementation we simulate a square channel as in 
Fig.~\ref{fig:TM}. The FEM shell on top of the channel reproduces an EPDM-rubber as in Sec.~\ref{sec:flag}, the channel dimensions are $65 \times 65 \times 65$cm, with a fluid density of $1000$kg/m$^3$, and with simulation resolution of $65 \times 65 \times 65$ cells. A pressure gradient $dP/dx \approx 30$Pa/m is applied across the flow direction ($x$) which results to a steady state flow of Re $\approx 1$. This flow resembles a Poiseuille profile flow through a cylinder, which is expected as a square channels can be visualized as an approximation to a cylindrical tube. When steady state is reached, an external, uniform pressure is applied at the top (0.37MPa). This pressure causes the membrane to bend in between its pinned boundaries, i.e. the sides of the top square, affecting the flow through the channel. 

As it can be seen from  Fig.~\ref{fig:TM}, a transverse velocity component ($v_z$) naturally develops from the motion and deformation of the wall, thus affecting the streamwise velocity but not disturbing it completely. Additionally there are some minor $y$ and $x$ velocity components created by the displacement of the fluid as the membrane moves down, these can be inferred by following the streamlines of Fig.~\ref{fig:TM}. 

\begin{figure}
\includegraphics[width=1.0\columnwidth]{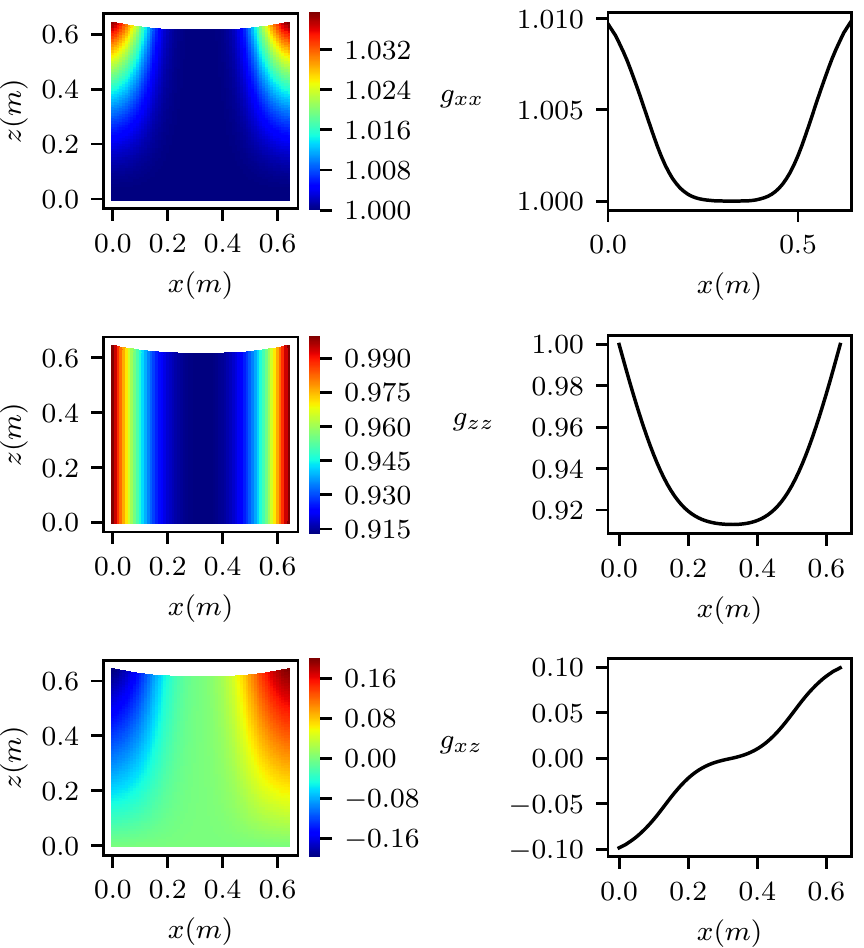}
\caption{\textbf{Metric Tensor.} Three $g_{ij}$ components shown on the $xz$ slice (left) and along the x direction for $z=L_z/2$ (right). } 
\label{fig:TM_metric}
\end{figure}

Furthermore the components of the metric are monitored, see Fig.~\ref{fig:TM_metric}. As explained in Sec.~\ref{sec:interpolation}, a smooth metric variation is required such as to ensure accurate calculation of the forcing term (Eqs.~\ref{eq:forcing-A},\ref{eq:forcing-B},\ref{eq:forcing-C}). This is achieved here by using the exterior boundary interpolation as explained in Sec.~\ref{sec:interpolation}. The LBM manifold is compressed along the $z$-direction as the membrane moves downwards.

\begin{figure}
\includegraphics[width=1.0\columnwidth]{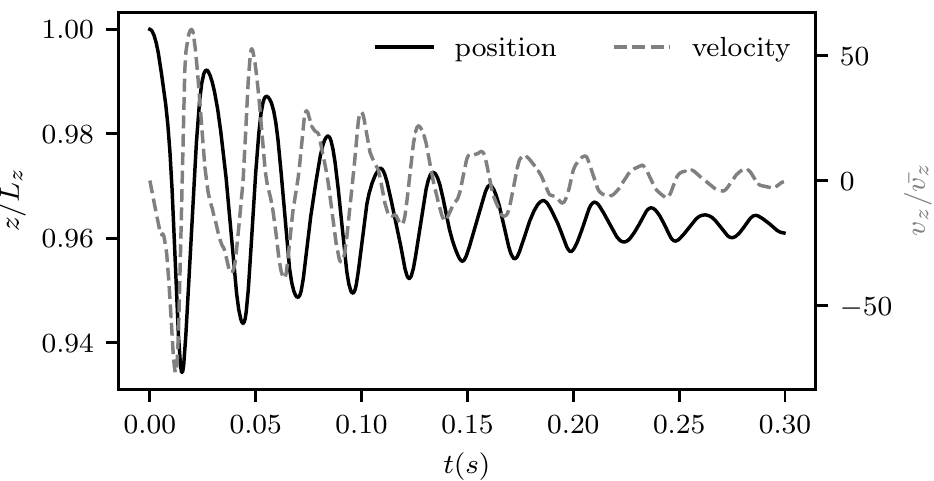}
\caption{ \textbf{Position and velocity of membrane.} Central node's position and velocity of FEM shell is tracked along the simulation. Position and velocity axes are normalized by the total distance and average velocity respectively.}
\label{fig:TM_oscillation}
\end{figure}

As seen in Fig.~\ref{fig:TM_oscillation}, by measuring the position and velocities at the center of the membrane we observe a convergence of the motion to an equilibrium position. The depth of the equilibrium configuration, naturally, depends on the pressure applied, the material, the velocity and the density of the fluid. Furthermore, we observe how the velocity oscillates with $\pi/2$ out of phase relative to the position. 

To set a measure of the gain in computational efficiency, the total memory allocation of the current LBM approach is around $45 \%$  and the total simulation time is around $70 \%$ when these are compared to the original curved space LBM. Additionally, within the complete coupled simulation, updating the chart $\mathbf{h}$ (including the interpolations) uses $6 \%$ of the total time, where transforming the distribution function and calculating the forcing term use $40 \%$.  These are equivalent to only $51 \%$ of the total LBM time. Therefore, the method is efficient even relative to a standard LBM solver.
\nomenclature{$L_i$}{Simulation domain size, i=x,y,z}%

\section{Summary and Conclusions}
We presented a novel method for fluid structure interaction simulations where differential geometry is used as the motivation for implementing deformable boundaries. Moreover, utilizing direct force and momentum calculation from the fluid solver, a precise coupling is achieved between the curved space LBM and the FEM shell. The curved space LBM uses a standard lattice, avoiding the need for off-lattice interpolation.

We demonstrated that our 2$^{nd}$-order curved space lattice Boltzmann reproduces both a Poiseuille profile under a metric deformation and a Taylor-Green vortex under a dynamic metric deformation, with resolution convergent errors. The volume conservation and pressure maintenance were verified by a closed deformed membrane scenario. In addition, boundary slip was validated through the laminar circular Couette flow. A brief stability analysis was performed, which indicated validity of the method within $10\%$ deformations relative to domain size. The restriction to small deformations mainly originates from the metric variation and is partly influenced by the reduction of the lattice speeds, which may introduce higher order anisotropies into the forcing term.  Additionally, we presented two applications of the method one at low and one at medium Reynolds numbers.

At low Reynolds number a deformable square channel has produced good qualitative results. The flag scenario (rubber in air at Re $\approx 230$) exhibited the expected physical behavior, when light damping was implemented, with a steady oscillation solution.
The spatial flag shape and the measured frequency  $\approx 5.1$ agreed, within error, to similar simulation results.

Future work can include direct, quantitative comparison of the flag oscillation with experimental and other numerical results. Additionally, an extended 3-dimensional flag can be simulated which is a more challenging example. Other possibilities could include different materials and fluids. Similarly the square channel can be extended to a pipe with deformable walls. 
The model stability, can be improved with an entropic LBM  \cite{entropic_lbm}, using Multi-relaxation-time LBM \cite{MRT_lbm} or  using a larger lattice as in higher-order isotropic LBMs \cite{high_order_lbm}. The method can then be used to simulate strongly deformed confining surfaces where the deformation can come about various factors. This opens up the possibility of studying complicated morphologies with high local curvatures, created by the fluid flow on a surface wall,  bringing a new perspective to biological morhogenesis and biomorphic technologies.

\printnomenclature

\appendix
\section{Riemannian geometry}
\label{app:riemannian}

In all appendices the Latin indices run over the spatial dimensions and Einstein summation convection is used for repeated indices.

A $D$ dimensional curved space is represented by a Riemannian manifold M, which is locally described by a smooth diffeomorphism $\mathbf{h}$, called the chart. The set of tangential vectors attached to each point $\mathbf{y}$ on the manifold is called the  tangent space $T_\mathbf{y} M$. In the fluid model, all the vector quantities are represented as elements of $T_\mathbf{y} M$. The derivatives of the chart $\mathbf{h}$ are used to define the standard basis $(\textbf{e}_1,...,\textbf{e}_D)=\frac{\del\mathbf{h}}{\del x^1},...,\frac{\del \mathbf{h}}{\del x^D}$. 

The metric tensor $g$, acting as a generalized dot product, can be used to measure the length of a vector or the angle between two vectors. In local coordinates, the components of the metric tensor are given by 
\begin{equation}
g_{ij}(x)= \textbf{e}_i(x)\cdot \textbf{e}_j(x)= \frac{\del \mathbf{h}}{\del x^i} \cdot \frac{\del \mathbf{h}}{\del x^j},
\end{equation}
where $\cdot$ is the standard Euclidean scalar product.

For a given metric tensor, the vector $v=v^i\textbf{e}_i \in T_\mathbf{y} M$ has a norm $||v||_g=\sqrt{v^ig_{ij}v^j}$ and a corresponding dual vector $v^*=v^i\textbf{e}_i \in T^*_\mathbf{y} M$ in the cotangent space, which is spanned by the differential 1-forms $dx^i=g(\textbf{e}_i,\cdot)$. The coefficients $v_i$ of the dual vector are typically denoted by a lower index and are related to the upper-index coefficients $v^i$ by contraction with the metric tensor $v_i = g_{ij}v^j$ or equivalently, $ v^i=g^{ij}v_j$, where $g^{ij}$ denotes the inverse of the metric tensor. The upper-index coefficients $v^i$ of a vector $v$ are typically called  \textit{contravariant components}, whereas the lower-index coefficients $v_i$ of the dual vectors $v^*$ are known as the \textit{covariant components}.

A necessary feature for the description of objects moving on the manifold is parallel transport of vectors along the manifold.  The tangent space is equipped with a covariant derivative $\nabla$ (Levi-Civita connection), which connects the tangent spaces at different points on the manifold and thus allows to transport a tangent vector  from one tangent space to the other along a given curve  $\gamma(t)$. The covariant derivative can be viewed  as the orthogonal projection of the Euclidean derivative $\del$ onto the tangent space, such that the tangency of the vectors is preserved during the transport. In local coordinates, the covariant derivative is fully characterized by its connection coefficients $\Gamma^i_{jk}$  (The Christoffel symbols), which are defined by the action of the covariant derivative on the basis vector, $\nabla_j \textbf{e}_k= \Gamma^i_{jk}$. In the standard basis, $\textbf{e}_i = \frac{\del \mathbf{h}}{\del x^i}$, the Christoffel symbols are related to the metric by
\begin{equation}
\Gamma^i_{jk}=\frac{1}{2}g^{ij}(\del_j g_{kl} + \del_k g_{jl} - \del_l g_{jk}).
\end{equation} 
Acting on a general vector $v=v^i \mathbf{e}_i,$ the covariant derivative becomes:
 \begin{equation}
\nabla_k v =(\del_k v^i + \Gamma^i_{kj}v^j)\mathbf{e}_i,
\end{equation}
where the product rule has been applied, using that the covariant derivative acts as a normal derivative on the scalar functions 
$v^i$. Extending to tensors of higher rank, for example the second order tensors $T= T^{ij} $, 
\begin{equation}
\nabla_k T=( \del_kT^{ij}+ \G ^i _{kl} T^{lj} + \G^j_{kl} T^{il})\mathbf{e}_i \otimes \mathbf{e}_j
\end{equation}
in this work the basis vectors $\mathbf{e}_i $ are generally dropped. Compatibility of the covariant derivative with the metric tensor implies that $\nabla_k g^{ij}=\nabla_k g_{ij} =0$. This property allows us to commute the covariant derivative with the metric tensor for the raising or lowering of tensor indices in derivative expressions.

The motion of the particle can be described by the curve $ \gamma(t)$, which parametrizes the position of the particle at time $t$. The geodesic equation,  $ \nabla_{\dot{\gamma}} \dot{\gamma} =0 $, in local coordinates $\gamma(t)=\gamma^i(t)\mathbf{e}_i$ is defined by
\begin{equation}
\label{eq:geodesic}
\ddot{\gamma}^i + \Gamma_{jk}^i \dot{\gamma^j} \dot{\gamma^k} = 0. 
\end{equation}
The geodesic equation can be interpreted as the generalization of Newtons law of inertia to curved space. The solutions of Eq.~(\ref{eq:geodesic}) represent lines of constant kinetic energy on the manifold, i.e. the geodesics. 
	The Riemann curvature tensor $R$ can be used to measure curvature, or more precisely, it measures curvature-induced change of a tangent vector $v$ when transported along a closed loop.
 \begin{equation}
R(\textbf{e}_i,\textbf{e}_j)v=\nabla_i \nabla_j v-\nabla_j \nabla_i v.  
\end{equation}   
In a local coordinate basis $ { \textbf{e}_i } $, the coefficients of the Riemann curvature tensor are given by
\begin{multline}
R^l_{ijk}= g(R(\textbf{e}_i,\textbf{e}_j)\textbf{e}_k,\textbf{e}_l) =
\\
\del_j \Gamma^l_{ik} - \del_k \Gamma^l_{ij} + \Gamma^l_{jm} \Gamma^m_{ik} -\Gamma^l_{km} \Gamma^m_{ij}.
\end{multline}
Contraction of $R^i_{jkl}$ to a rank 2 and 1 tensor yields  the Ricci-tensor $R_{ij}=R^k_{ikj}$ and the Ricci-scalar $R=g^{ij}R_{ij}$ respectively, which can also be used to quantify curvature.

Finally some general definitions of operators in curved space. The gradient $\nabla^i f= g^{ij} \del_j f$, divergence $\nabla_i v^i= \frac{1}{\sqrt{g}} \del_i (\sqrt{g} v^i) $, integration over curved volume $V=\int_V QdV$, where $dV=\sqrt{g}dx^1...dx^D=:\sqrt{g}d^Dx$ denotes the volume element. $\sqrt{g}$ denotes the square root of the determinant of the metric tensor.   

It should be clarified that in the simulations there is no time curvature and $g_{ij}$ denotes the curved space metric not space-time.

\section{\label{sec:appendix_Hermite} Gauss Hermite quadrature}
 The expansion coefficients of the equilibrium distribution are given by:
\begin{align}\label{eq:hermite-coefficients-eq}
	a_{(0)}^{{\rm eq}} &= \r,\\
	a_{(1)}^{{\rm eq},i} &= \r u^i,\\ 
	a_{(2)}^{{\rm eq},ij} &= \r (\theta g^{ij} - c_s^2 \delta^{ij}) + \r u^i u^j ,
\end{align}
where $\theta$ denotes the normalized temperature (from the Maxwell-Boltzmann equilibrium distribution), $c_s$ the lattice-specific speed of sound and $\delta^{ij}$ is the Kronecker delta.

Using an expanded polynomial basis we can solve the Boltzmann equation numerically, by expanding the distribution function and the forcing term. We use a polynomial basis that is accurately reproducing the Gaussian shaped Maxwell Boltzmann equilibrium distribution. This is the case for the Hermite polynomials defined as such:
\begin{equation}
{\cal H}^{i_1....i_n}_{(n)} (v) = (-1)^n w(v)^{-1} \frac{\del}{\del v_{i1}}..... \frac{\del}{\del v_{in}}w(v)
\end{equation}
where $w(v)$ is the weight function, is given by
\begin{equation}
w(v)=\frac{1}{(2*\pi)^{d/2}}exp(-\frac{1}{2}||v||^2)^2
\end{equation} 
The Hermite polynomials are given by
\begin{align}
	\HH_{(0),\l} &= 1, \\
	\HH_{(1),\l}^i &= \frac{c_\l^i}{c_s}, \\
	\HH_{(2),\l}^{ij} &= \frac{c_\l^i c_\l^j}{c_s^2} - \d^{ij}, \\	
	\HH_{(3),\l}^{ij} &= \frac{c_\l^i c_\l^j c_\l^k}{c_s^3} - \d^{ij} \frac{c_\l^k}{c_s} - \d^{jk} \frac{c_\l^i}{c_s} - \d^{ik} \frac{c_\l^j}{c_s} \\
	&\vdots
\end{align}
Thus, the equilibrium distribution satisfies the following relations:
\begin{align}
	\r 	&= \sum_\l f^{\rm eq}_\l \,\sqrt g, \\
	\r u^i	&= \sum_\l f^{\rm eq}_\l c_\l^i \,\sqrt g, \\
	\label{eq:feq-pi}
	\Pi^{{\rm eq},ij}	&= \sum_\l f^{\rm eq}_\l c_\l^i c_\l^j \,\sqrt g 	= \r\left( \theta g^{ij} + u^i u^j \right).
\end{align}

As seen by plugging in the explicit expressions of the first Hermite polynomials the coefficients $a_{(n)}$  can recover the macroscopic moments of the distribution function.
 With this LB method the calculation of the macroscopic moments on the lattice is exact under the use of Gauss-Hermite quadrature, this implies that the integrals over velocity space are exchanged to sums on a discrete set without loss of accuracy. To satisfy the orthogonality relations to N order the lattice is required to obey certain symmetries.

\section{Chapman-Enskog expansion}
\label{app:chapmanenskog}

So as to prove that the LB equation converges to the Navier-Stokes equations in the hydrodynamic limit of small Knudsen numbers, we perform a Chapman-Enskog multiscale analysis. To this end, we firstly perform a Taylor expansion of the distribution function in Eq. (\ref{eq:CE-LB}), which yields:
\begin{multline}\label{eq:CE-LB2}
	\dt\, D_t f_\l + \frac{\dt^2}{2!} D_t^2 f_\l + \ldots 
	= - \frac{1}{\t} \left( f_\l - f_\l^{\rm eq} \right) + \\
	 + \dt\, \FF_\l + \frac{\dt^2}{2} D_t \FF_\l + \ldots,
\end{multline}
where $D_t := \del_t + c_\l^i \del_i$ denotes the material derivative, and the dots indicate irrelevant higher order terms $\sim \OO(\dt^3)$.
The distribution function as well as the time and space derivatives are now expanded in terms of the Knudsen number $\e$:
\begin{gather*}
	f = f^{(0)} + \e f^{(1)} + \e^2 f^{(2)} + ... , 
	\qquad
	\del_t = \e \del_t^{(1)} + \e^2 \del_t^{(2)} + ... , \\
	(\del_i,\FF, A, B^i, C^{ij}) = \e (\del_i^{(1)}, \FF^{(1)}, A^{(1)}, B^{(1),i}, C^{(1),ij}).
\end{gather*}
Plugging everything into Eq. (\ref{eq:CE-LB2}) and comparing orders of $\e$, we obtain the following set of equations:
\begin{align}
	\label{eq:CE-0}
	&\T\OO(\e^0): \quad
	f_\l^{(0)} = f_\l^{{\rm eq}}, \\ 
	\label{eq:CE-I}
	&\T\OO(\e^1): \quad
	\T D_t^{(1)} f_\l^{(0)} = - \frac{1}{\t \dt} f_\l^{(1)} + \FF_\l^{(1)},\\	
	\label{eq:CE-II}
	&\T\OO(\e^2): \quad
	\T\del_t^{(2)} f_\l^{(0)} + \left(1- \frac{1}{2\t}\right) D_t^{(1)} f_\l^{(1)}
	\T= - \frac{1}{\t \dt} f_\l^{(2)}.
\end{align}
The moments of $f_\l^{(0)} = f_\l^{\rm eq}$ and $f_\l^{(1)}$ can be deduced from the fact that the collision operator conserves mass and momentum, i.e. 
\begin{multline}
  \sum_\l f_\l \,\sqrt g
  = \sum_\l f_\l^{\rm eq} \,\sqrt g
  = \r,
\\
 \sum_\l f_\l c_\l^i \,\sqrt g
  = \sum_\l f_\l^{\rm eq} c_\l^i \,\sqrt g
  = \r u^i.
\end{multline}
Thus we find:\\
\begin{minipage}{0.4\textwidth} 
	\begin{align}
		&\sum_\l f_\l^{(0)} \,\sqrt g= \r,\\
		&\sum_\l f_\l^{(0)} c_\l^i \,\sqrt g = \r u^i,\\
		&\sum_\l f_\l^{(0)} c_\l^i c_\l^j \,\sqrt g 
		= \Pi^{(0),ij} = \Pi^{{\rm eq},ij},
	\end{align}
\end{minipage}
\hfill
\begin{minipage}{0.4\textwidth}
	\begin{align}
		&\sum_\l f_\l^{(1)} \,\sqrt g= 0,\\
		&\sum_\l f_\l^{(1)} c_\l^i \,\sqrt g = 0,\\
		&\sum_\l f_\l^{(1)} c_\l^i c_\l^j \,\sqrt g 
		= \Pi^{(1),ij},
	\end{align}
\end{minipage}\\
where $\Pi^{{\rm eq},ij} = \r\left(\theta g^{ij} + u^i u^j \right) $.

\subsection*{Moments of Eq. (\ref{eq:CE-I}-\ref{eq:CE-II})}

Taking the moments of Eq. (\ref{eq:CE-I}) yields:
\begin{align}
	\label{eq:CE-1}
	\T\sum_\l (\ref{eq:CE-I}) \sqrt g :
	\T\quad &\del_t^{(1)} \r + \delvar_i^{(1)} \left(\r u^i \right)
	\T= A^{(1)}, \\
	\label{eq:CE-2}
	\T\sum_\l c_\l^i\, (\ref{eq:CE-I}) \sqrt g  : \quad &
	\T\del_t^{(1)} \left( \r u^i \right) + \delvar_j^{(1)} \Pi^{(0),ij}
	\T= B^{(1),i},
\end{align}
where $A$ and $B^i$ are the moments of the forcing term (\ref{eq:forcing-A}-\ref{eq:forcing-B}) and $\delvar_i := \del_i - \G^j_{ij}$. Here, the additional Christoffel symbol term in the derivative $\delvar_i$ originates from the metric determinant $\sqrt g$:
\begin{multline}
	\T\sum_\l c_\l^i \del_i f_\l^{(0)} \sqrt g
	= \T\left(\del_i - \frac{\del_i \sqrt g}{\sqrt g}\right)
	\sum_\l c_\l^i f_\l^{(0)} \sqrt g \\
	= \del_i (\r u^i) - \G^j_{ij} (\r u^i) 
	=: \delvar_i (\r u^i),
\end{multline}
where we have used the identity $\del_i \sqrt g = \G^j_{ij} \sqrt g$.
The moments of Eq. (\ref{eq:CE-II}) are given by
\begin{align}
	\label{eq:CE-3}	
	\T\sum_\l (\ref{eq:CE-II}) \sqrt g : \quad&
	\T\del_t^{(2)} \r = 0, \\	
	\label{eq:CE-4}
	\T\sum_\l c_\l^i\, (\ref{eq:CE-II}) \sqrt g : \quad&
	\T\del_t^{(2)} \left(\r u^i \right) 
	\T= \delvar_j^{(1)} \s^{(1),ij},
\end{align}
where $\s^{(1),ij}$, the viscous stress tensor (rescaled by $\e$), is given by
\begin{align}\label{eq:CE-viscous-stress-tensor}
	\T\s^{(1),ij}
	= - \left(1 - \frac{1}{2\t}\right) \Pi^{(1),ij}
	= - \left(1 - \frac{1}{2\t}\right) \sum_\l  f^{(1)}_\l c_\l^i c_\l^j\, \sqrt g.
\end{align}

\subsection*{Continuity Equation}

For the continuity equation, we add $\e \cdot (\ref{eq:CE-1})$ and $\e^2 \cdot (\ref{eq:CE-3})$:
\begin{align}
	\T\quad\del_t \r + \delvar_i \left(\r u^i \right)
	= A.
\end{align}
After inserting the explicit expression for $A$ (\ref{eq:forcing-A}), we obtain the continuity equation with error $\OO(\dt^2)$:
\begin{align}\label{eq:CE-continuity}
	\del_t \r + \cov_i \left(\r u^i \right) = 0 + \OO(\dt^2)
\end{align}
where $\cov$ denotes the covariant derivative.

\subsection*{Momentum Equation}

Adding $\e \cdot(\ref{eq:CE-2})$ and $\e^2 \cdot (\ref{eq:CE-4})$ yields the momentum conservation equation:
\begin{align}
	\T \del_t \left( \r u^i \right) 
	+ \delvar_j \Pi^{(0),ij}
	\T= \delvar_j \s^{ij} + B^i.
\end{align}
Inserting the explicit expressions for $B^i$ (\ref{eq:forcing-B}) and $\Pi^{(0),ij}$ (\ref{eq:feq-pi}) yields the Navier-Stokes momentum equation with error $\OO(\dt^2)$:
 \begin{align}\label{eq:CE-momentum}
	\del_t \left( \r\, u^i \right) + \cov_j \left( \r\, u^i u^j + \r \,\theta g^{ij} \right) 
	&= \T \cov_j \s^{ij} + \OO(\dt^2),
\end{align}
where $\s^{ij}$ is the viscous stress tensor, whose explicit form in terms of $\r$ and $u^i$ will be derived in the next section.

\subsection*{Viscous Stress Tensor}

For the derivation of the viscous stress tensor $\s^{ij}$, we rewrite
\begin{align}
	\s^{ij} \overset{(\ref{eq:CE-viscous-stress-tensor})}=& \T -\left( 1 - \frac{1}{2\t}\right) 
	\e \sum_\l c_\l^i c_\l^j f_\l^{(1)} \sqrt g \\
	\overset{(\ref{eq:CE-I})}=& \T \left( \t - \frac{1}{2}\right) \dt\, \e\, \sum_\l c_\l^i c_\l^j  \left( D_t^{(1)} f_\l^{(0)} - \FF_\l^{(1)} \right) \sqrt g \\
	\approx\ \ & \T \left( \t - \frac{1}{2}\right) \dt \Big( \delvar_k (c_s^2 \r (u^i \d^{jk} + u^j \d^{ik} + u^k \d^{ij})) - C^{ij} \Big).
\end{align}
Here, we have used
\begin{align*}
	&\sum_\l c_\l^i c_\l^j \del_t^{(1)} f_\l^{(0)} \,\sqrt g 
	\approx 0, \\
	&\sum_\l c_\l^i c_\l^j c_\l^k \del_k^{(1)} f_\l^{(0)} \,\sqrt g 
	= \delvar_k (c_s^2 \r (u^i \d^{jk} + u^j \d^{ik} + u^k \d^{ij})).
\end{align*}
After plugging in the explicit expression for $C^{ij}$ (\ref{eq:forcing-C}), we obtain
\begin{align}\label{eq:CE-viscous-stress-tensor2}
	\s^{ij} =  \T \nu \left( \cov^j (\r u^i) + \cov^i (\r u^j)  + g^{ij} \cov_k (\r u^k) \right),
\end{align}
where we have defined $\nu := \left( \t - \frac{1}{2}\right) \dt\, \theta$ and neglected terms of the order $\OO(u^3) \sim \OO(\text{Ma}^3)$, $\text{Ma}$ being the Mach number.


 \section{Improved Lattice derivatives}
 \label{app:derivative}
\begin{figure}
\includegraphics[width=0.5\columnwidth]{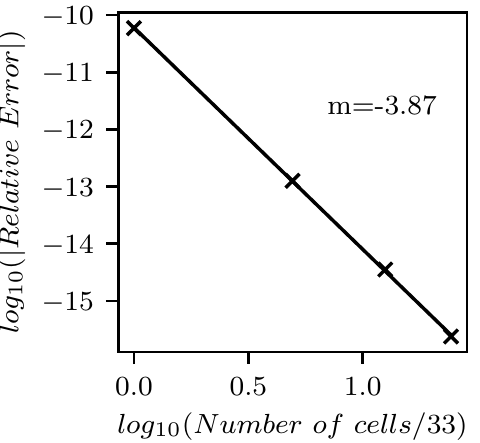}
\caption{\label{fig:nabla_g} \textbf{Derivative error convergence.} The relative error of the metric tensor derivative is plotted against increasing resolution, $m$ denotes the gradient.}
\end{figure}

Here we present an improved way of calculating discrete derivatives on the lattice with up to forth-order precision. We use the isotropic discrete differential operators as proposed in \cite{succi13_ld}. The gradient and the Laplacian of a lattice function $f$ are given by
\begin{align}
& \ \ \  \del_i f(x)= \frac{1}{c^2_s \Delta t} \sum_\l w_\l c_\l^i f(x+c_\l \Delta t) + \mathcal{O}(\Delta t^2), \\
& \ \ \  \Delta f(x) = \frac{2}{c^2_s \Delta t} \sum_\l w_\l ( f(x+c_\l \Delta t)-f(x)) + \mathcal{O}(\Delta t^2) \label{eq:der2},
\end{align}
using the lattice symmetries is furthermore possible to calculate the derivative to 4th order with respect to $ \Delta t$;
\begin{multline}
 \del_i f(x)= \frac{1}{c^2_s \Delta t} \sum_\l w_\l c_\l^i ( \Delta f(x+c_\l \Delta t) \notag\ \\ - \frac{c^2_s \D t^2}{x}(\D f)(x+c_\l \Delta t)   ) + \mathcal{O}(\D t^4), 
\end{multline}
where $\Delta f$ denotes the Laplacian of $f$, which can precomputed using Eq.~(\ref{eq:der2}). Therefore all the differential geometrical object such as the metric, the Christoffel symbols and the Ricci curvature can be precomputed at high numerical precision $\mathcal{O}(\D t^4)$. In this work the 19 second nearest nodes, as in Fig.~\ref{fig:lb_velocities}, are used to calculate the derivatives. The precision is investigated by comparing the error of a sinusoidal metric tensor derivative to the analytical solution, see Fig.~\ref{fig:nabla_g}.

\begin{acknowledgments}
The authors are grateful for the financial support of the ETH Zurich under Grant No. 06 11-1, and the European Research Council (ERC) Advanced Grant 319968-FlowCCS.
\end{acknowledgments}
\bibliographystyle{ieeetr}


\begin{thebibliography}{10}

\bibitem{fsi_review_heart}
I.~Borazjani, ``A review of fluid--structure interaction simulations of
  prosthetic heart valves,'' {\em Journal of Long-Term Effects of Medical
  Implants}, vol.~25, no.~1-2, pp.~75--93, 2015.

\bibitem{review_heart2}
B.~T. Chan, E.~Lim, K.~H. Chee, and N.~A.~A. Osman, ``Review on {CFD}
  simulation in heart with dilated cardiomyopathy and myocardial infarction,''
  {\em Computers in Biology and Medicine}, vol.~43, no.~4, pp.~377 -- 385,
  2013.

\bibitem{fsi_book}
T.~E.~T. Yuri~Bazilevs, Kenji~Takizawa, {\em Computational Fluid-Structure
  Interaction: Methods Applications}.
\newblock Wilet, 2013.

\bibitem{high_local_curvature}
T.~Baumgart, S.~T. Hess, and W.~W. Webb, ``Imaging coexisting fluid domains in
  biomembrane models coupling curvature and line tension,'' {\em Nature},
  vol.~425, p.~821, 2003.

\bibitem{temp_control_fluids}
V.~Miralles, A.~Huerre, F.~Malloggi, and M.-C. Jullien, ``A review of heating
  and temperature control in microfluidic systems: Techniques and
  applications,'' {\em Diagnostics}, vol.~3, no.~1, pp.~33--67, 2013.

\bibitem{surface_chemistry}
G.~A. Somorjai and Y.~Li, ``Impact of surface chemistry,'' {\em Proceedings of
  the National Academy of Sciences}, vol.~108, no.~3, pp.~917--924, 2011.

\bibitem{FSI_engine}
W.-B. Shangguan and Z.-H. Lu, ``Experimental study and simulation of a
  hydraulic engine mount with fully coupled fluid--structure interaction
  {Finite Element} analysis model,'' {\em Computers \& Structures}, vol.~82,
  no.~22, pp.~1751--1771, 2004.

\bibitem{FSI_heart}
C.~Yang, D.~Tang, N.~Vasilyev, R.~Rathod, T.~Geva, and P.~J. del Nido, ``{FSI}
  modeling approach to develop right ventricle pulmonary valve replacement
  surgical procedures with a contracting actuator and improve ventricle
  ejection fraction,'' {\em Procedia Engineering}, vol.~126, pp.~441--445,
  2015.

\bibitem{FSI_vesels}
D.~Bluestein, Y.~Alemu, I.~Avrahami, M.~Gharib, K.~Dumont, J.~J. Ricotta, and
  S.~Einav, ``{Influence of microcalcifications on vulnerable plaque mechanics
  using {FSI} modeling },'' {\em Journal of Biomechanics}, vol.~41, no.~5,
  pp.~1111--1118, 2008.

\bibitem{finite_elements_book}
K.~J. Bathe, {\em Finite element procedures}.
\newblock No.~9780979004902, Prentice Hall, 2006.

\bibitem{review_fsi_types}
G.~Hou, J.~Wang, and A.~Layton, ``Numerical methods for fluid-structure
  interaction — a review,'' {\em {Communications in Computational Physics}},
  vol.~12, no.~2, pp.~337--377, 2012.

\bibitem{immersed}
F.~Sotiropoulos and X.~Yang, ``Immersed boundary methods for simulating
  fluid–structure interaction,'' {\em Progress in Aerospace Sciences},
  vol.~65, no.~Supplement C, pp.~1 -- 21, 2014.

\bibitem{immersed2}
I.~Borazjani, ``Fluid–structure interaction, immersed boundary-finite element
  method simulations of bio-prosthetic heart valves,'' {\em Computer Methods in
  Applied Mechanics and Engineering}, vol.~257, no.~Supplement C, pp.~103 --
  116, 2013.

\bibitem{fsi_lbm_fem1}
A.~D. Rosis, G.~Falcucci, S.~Ubertini, and F.~Ubertini, ``A coupled lattice
  boltzmann-finite element approach for two-dimensional fluid–structure
  interaction,'' {\em Computers \& Fluids}, vol.~86, pp.~558 -- 568, 2013.

\bibitem{fsi_lbm_fem2}
Y.~W. Kwon, ``Coupling of lattice boltzmann and finite element methods for
  fluid-structure interaction,'' {\em Journal of Pressure Vessel Technology},
  vol.~130, pp.~011302--011302--6, 2008.

\bibitem{fsi_lbm_fem3}
M.~Garcia, J.~Gutierrez, and N.~Rueda, ``Fluid-structure coupling using
  lattice-boltzmann and fixed-grid fem,'' {\em Finite Elem. Anal. Des.},
  vol.~47, no.~8, pp.~906--912, 2011.

\bibitem{LBM_review}
S.~Chen and G.~D. Doolen, ``Lattice boltzmann method for fluid flows,'' {\em
  Annual Review of Fluid Mechanics}, vol.~30, no.~1, pp.~329--364, 1998.

\bibitem{LBM_initial}
X.~He and L.-S. Luo, ``Theory of the lattice {Boltzmann} method: From the
  {Boltzmann} equation to the lattice {Boltzmann} equation,'' {\em Phys. Rev.
  E}, vol.~56, pp.~6811--6817, Dec 1997.

\bibitem{miller_lks}
M.~Mendoza, J.~Debus, S.~Succi, and H.~Herrmann, ``{L}attice kinetic scheme for
  generalized coordinates and curved spaces,'' {\em International Journal of
  modern physics C, Computational physics, physical computation}, vol.~25,
  no.~12, p.~1441001, 2014.

\bibitem{shellcode}
R.~Vetter, N.~Stoop, T.~Jenni, F.~K. Wittel, and H.~J. Herrmann, ``Subdivision
  shell elements with anisotropic growth,'' {\em Journal for Numerical Methods
  in Engineering}, vol.~95, pp.~791--810, 2013.

\bibitem{roman_thesis}
R.~Vetter, ``Growth, interaction and packing of thin objects,'' {\em
  ETH-Z\"{u}rich}, 2015.

\bibitem{OL-finitevolume}
D.~V. Patil and K.~Lakshmisha, ``{Finite volume TVD formulation of lattice
  Boltzmann simulation on unstructured mesh},'' {\em Journal of Computational
  Physics}, vol.~228, no.~14, pp.~5262 -- 5279, 2009.

\bibitem{OL-finitevolume2}
F.~Nannelli and S.~Succi, ``The lattice boltzmann equation on irregular
  lattices,'' {\em Journal of Statistical Physics}, vol.~68, no.~3,
  pp.~401--407, 1992.

\bibitem{OL-finiteelement}
A.~Düster, L.~Demkowicz, and E.~Rank, ``High-order finite elements applied to
  the discrete {Boltzmann} equation,'' {\em {International Journal for
  Numerical Methods in Engineering}}, vol.~67, no.~8, pp.~1094--1121, 2006.

\bibitem{OL-finitediff}
A.~Fakhari and T.~Lee, ``Numerics of the lattice {Boltzmann} method on
  nonuniform grids: Standard {LBM} and finite-difference {LBM},'' {\em
  Computers \& Fluids}, vol.~107, no.~Supplement C, pp.~205 -- 213, 2015.

\bibitem{OL-finitediff2}
``Implementation of a high-order compact finite-difference lattice {Boltzmann}
  method in generalized curvilinear coordinates,'' {\em {Journal of
  Computational Physics}}, vol.~267, no.~Supplement C, pp.~28 -- 49, 2014.

\bibitem{lbm_review_succi}
S.~Succi, ``Lattice boltzmann 2038,'' {\em EPL (Europhysics Letters)},
  vol.~109, no.~5, p.~50001, 2015.

\bibitem{succi_qlbm}
S.~Succi and R.~Benzi, ``{Lattice Boltzmann equation for quantum mechanics},''
  {\em Physica D: Nonlinear Phenomena}, vol.~69, no.~3, pp.~327 -- 332, 1993.

\bibitem{miller_rlbm}
M.~Mendoza, B.~M. Boghosian, H.~J. Herrmann, and S.~Succi, ``Fast lattice
  {Boltzmann} solver for relativistic hydrodynamics,'' {\em Physical Review
  Letters}, vol.~105, p.~014502, 2010.

\bibitem{debus_curved}
J.-D. Debus, M.~Mendoza, S.~Succi, and H.~J. Herrmann, ``Poiseuille flow in
  curved spaces,'' {\em Physical Review E}, vol.~93, p.~043316, 2016.

\bibitem{miller_campylotic}
{Mendoza M.}, {Succi S.}, and {Herrmann H. J.}, ``{Flow Through Randomly Curved
  Manifolds},'' {\em Scientific Reports}, vol.~3, p.~3106, 2013.

\bibitem{discrete_lattice_effects}
Z.~Guo, C.~Zheng, and B.~Shi, ``Discrete lattice effects on the forcing term in
  the lattice boltzmann method,'' {\em Phys. Rev. E}, vol.~65, p.~046308, 2002.

\bibitem{Debus_thesis}
J.-D. Debus, {\em Flows in curved spaces}.
\newblock PhD thesis, ETH Zurich, 2016.

\bibitem{latt_2006}
J.~Latt and B.~Chopard, ``Lattice {Boltzmann} method with regularized
  pre-collision distribution functions,'' {\em Mathematics and computers in
  simulation}, vol.~72, no.~2-6, pp.~165--168, 2006.

\bibitem{cirak1}
F.~Cirak and M.~Ortiz, ``Fully c1-conforming subdivision elements for finite
  deformation thin-shell analysis,'' {\em International Journal for Numerical
  Methods in Engineering}, vol.~51, no.~7, pp.~813--833, 2001.

\bibitem{cirak2}
F.~Cirak, M.~Ortiz, and P.~Schröder, ``Subdivision surfaces: a new paradigm
  for thin-shell finite-element analysis,'' {\em International Journal for
  Numerical Methods in Engineering}, vol.~47, no.~12, pp.~2039--2072, 2000.

\bibitem{shellcode_thinsheets}
R.~Vetter, N.~Stoop, F.~K. Wittel, and H.~J. Herrmann, ``{Simulating Thin
  Sheets: Buckling, Wrinkling, Folding and Growth},'' {\em Journal of Physics:
  Conference Series}, vol.~487, no.~1, p.~012012, 2014.

\bibitem{loop}
C.~T. Loop, \em{Smooth Subdivision Surfaces Based on Triangles},
\newblock Masters Thesis, Department of Mathematics, The University of Utah, 1987.

\bibitem{stam}
J.~Stam, ``{Exact Evaluation of Catmull-Clark Subdivision Surfaces at Arbitrary
  Parameter Values},'' in {\em Proceedings of the 25th Annual Conference on
  Computer Graphics and Interactive Techniques}, {SIGGRAPH '98}, (New York, NY,
  USA), pp.~395--404, ACM, 1998.

\bibitem{roman_nature}
R.~Vetter, F.~K. Wittel, and H.~J. Herrmann, ``Morphogenesis of filaments
  growing in flexible confinements,'' {\em Nature Communications}, vol.~5,
  p.~4437, 2014.

\bibitem{Love}
A.~E.~H. Love, ``The small free vibrations and deformation of a thin elastic
  shell,'' {\em Philosophical Transactions of the Royal Society of London A:
  Mathematical, Physical and Engineering Sciences}, vol.~179, pp.~491--546,
  1888.

\bibitem{stability}
Y.~Cheng, L.~Zhu, and C.~Zhang, ``Numerical study of stability and accuracy of
  the immersed boundary method coupled to the lattice boltzmann bgk model,''
  {\em Communications in Computational Physics}, vol.~16, no.~1, p.~136–168,
  2014.

\bibitem{stability2}
Y.~Cheng, H.~Zhang, and C.~Liu, ``Immersed boundary-lattice boltzmann coupling
  scheme for fluid-structure interaction with flexible boundary,'' {\em
  Communications in Computational Physics}, vol.~9, no.~5, p.~1375–1396,
  2011.

\bibitem{stability3}
S.~Xu and Z.~J. Wang, ``An immersed interface method for simulating the
  interaction of a fluid with moving boundaries,'' {\em Journal of
  Computational Physics}, vol.~216, no.~2, pp.~454 -- 493, 2006.

\bibitem{flag_exp}
M.~J. Shelley and J.~Zhang, ``Flapping and bending bodies interacting with
  fluid flows,'' {\em Annual Review of Fluid Mechanics}, vol.~43, no.~1,
  pp.~449--465, 2011.

\bibitem{Turek2006}
S.~Turek and J.~Hron, {\em Proposal for Numerical Benchmarking of
  Fluid-Structure Interaction between an Elastic Object and Laminar
  Incompressible Flow}, pp.~371--385.
\newblock Berlin, Heidelberg: Springer Berlin Heidelberg, 2006.

\bibitem{Gilmanov2015}
A.~Gilmanov, T.~B. Le, and F.~Sotiropoulos, ``A numerical approach for
  simulating fluid structure interaction of flexible thin shells undergoing
  arbitrarily large deformations in complex domains,'' {\em Journal of
  Computational Physics}, vol.~300, pp.~814 -- 843, 2015.

\bibitem{entropic_lbm}
S.~Chikatamarla and I.~Karlin, ``Entropic lattice boltzmann method for
  turbulent flow simulations: Boundary conditions,'' {\em Physica A:
  Statistical Mechanics and its Applications}, vol.~392, no.~9, pp.~1925 --
  1930, 2013.

\bibitem{MRT_lbm}
R.~Du, B.~Shi, and X.~Chen, ``Multi-relaxation-time lattice boltzmann model for
  incompressible flow,'' {\em Physics Letters A}, vol.~359, no.~6, pp.~564 --
  572, 2006.

\bibitem{high_order_lbm}
F.~Dubois and P.~Lallemand, ``Towards higher order lattice boltzmann schemes,''
  {\em Journal of Statistical Mechanics: Theory and Experiment}, vol.~2009,
  no.~06, p.~P06006, 2009.

\bibitem{succi13_ld}
S.~P. Thampi, S.~Ansumali, R.~Adhikari, and S.~Succi, ``Isotropic discrete
  {Laplacian} operators from lattice hydrodynamics,'' {\em Journal of
  Computational Physics}, vol.~234, pp.~1 -- 7, 2013.

\end{thebibliography}


\end{document}